\documentclass[11pt,oneside,letterpaper]{article}
\usepackage[utf8]{inputenc}
\usepackage{graphicx}
\usepackage{amsmath}
\usepackage{amsthm}
\usepackage{amssymb}
\usepackage{bbm}
\usepackage{physics}
\usepackage{tensor}
\usepackage{slashed}
\usepackage[a4paper, total={6.6in, 9in}]{geometry}
\usepackage{mathrsfs}
\usepackage[linktocpage]{hyperref}
\usepackage{setspace}
\usepackage{comment}
\usepackage{tikz}
\usepackage{array}
\usepackage{longtable}
\usepackage{mathtools}
\usepackage{afterpage}
\usepackage{authblk}
\usepackage{cite}

\newcommand{\be}{\begin{equation}}
\newcommand{\ee}{\end{equation}}
\newcommand{\bea}{\begin{eqnarray}}
\newcommand{\eea}{\end{eqnarray}}
\newcommand{\bel}{\begin{align}}
\newcommand{\eel}{\end{align}}
\newcommand{\bi}{\begin{itemize}}
\newcommand{\ei}{\end{itemize}}
\newcommand{\id}{{\mathbbm{1}}}

\makeatletter\@addtoreset{equation}{section}\makeatother 

\newcommand\mc[1]{\mathcal{#1}}
\newcommand\mb[1]{\mathbb{#1}}

\definecolor{vert}{rgb}{0.1367 0.543 0.1367}

\hypersetup{
    colorlinks=true,
    citecolor=blue,
    linkcolor=blue,
    filecolor=magenta,      
    urlcolor=blue,
    pdfpagemode=FullScreen,
    }

\title{{\huge Resurgence in Liouville Theory} \vspace{1cm}}

\author[a, b]{Nathan Benjamin}
\author[c]{Scott Collier}
\author[d, e, f]{Alexander Maloney}
\author[d]{Viraj Meruliya\vspace{0.8cm}}

\affil[a]{\textit{{Walter Burke Institute for Theoretical Physics, Caltech, Pasadena, CA, 91125, USA
}}\vspace{0.2cm}}
\affil[b]{\emph{Department of Physics and Astronomy,
University of Southern California, \newline
Los Angeles, CA, 90089, USA
}\vspace{0.2cm}}
\affil[c]{\textit{{Center for Theoretical Physics, Massachusetts Institute of Technology, \newline Cambridge, MA, 02139, USA }\vspace{0.2cm}}
}
\affil[d]{\textit{{Department of Physics, McGill University, Montreal, QC, H3A 2T8, Canada }}\vspace{0.2cm}}
\affil[e]{\textit{{Department of Physics, Syracuse University, Syracuse, NY, 13244, USA}}\vspace{0.2cm}}
\affil[f]{\textit{{Institute for Quantum and Information Sciences, Syracuse University, \newline Syracuse, NY, 13244, USA}}}

\date{}

\begin{document}
   
\maketitle

\thispagestyle{empty}

\begin{center}
  \texttt{nathanbe@usc.edu, sac@mit.edu, admalone@syr.edu, viraj.meruliya@mail.mcgill.ca}
\end{center}
\vspace{1cm}

\begin{abstract}
Liouville conformal field theory is a prototypical example of an exactly solvable quantum field theory, in the sense that the correlation functions in an arbitrary background can be determined exactly using only the constraints of unitarity and crossing symmetry.  For example, the three point correlation functions are given by the famous formula of Dorn-Otto-Zamolodchikov-Zamolodchikov (DOZZ).  Unlike many other exactly solvable theories, Liouville theory has a continuously tunable parameter -- essentially $\hbar$ -- which is related to the central charge of the theory.  Here we investigate the nature of the perturbative expansion in powers of $\hbar$, which is the loop expansion around a semi-classical solution.  We show that the perturbative coefficients grow factorially, as expected of a Feynman diagram expansion, and take the form of an asymptotic series.
We identify the singularities in the Borel plane, and show that they are associated with complex instanton solutions of Liouville theory; they correspond precisely to the complex solutions described by Harlow, Maltz, and Witten.  Both single- and multi-valued solutions of Liouville appear. 
We show that the perturbative loop expansions around these different saddle points mix in the way expected for a trans-series expansion.  Thus Liouville theory provides a calculable example of a quantum field theory where perturbative and instanton contributions can be summed up and assembled into a finite answer.  

\end{abstract}

\newpage
\pagenumbering{arabic}
\tableofcontents

\section{Introduction}
In most quantum field theories the observables cannot be computed exactly, but only in a perturbative approximation.  One important example is the Feynman diagram expansion, which is a series expansion around a classical saddle point.
In most cases, however, this perturbative series is not convergent; it is typically an asymptotic series, and the addition of non-perturbative contributions are required to make sense of the theory \cite{Dyson:1952tj,Lipatov:1976ny,tHooft:1977xjm,shenker_strength_1991}. At the same time, non-perturbative contributions are often difficult to study directly.  If one could understand the detailed structure of the perturbative series -- in particular, if one can compute terms in the series to high enough order -- it is possible to recover information about the non-perturbative physics. This is the idea behind resurgence (see e.g. \cite{ecalle1985fonctions,Dorigoni:2014hea,Marino:2012zq,dingle1973asymptotic,83911f78-c315-3c02-ac25-bb06ed705713,f1d7d86e-5c82-323a-9807-a6be125c47d8,costin_asymptotics_2008,tournier_introduction_1994}), a beautiful tool that can be used to understand exactly how non-perturbative and perturbative contributions are related to one another. This has been successfully applied to a variety of systems, but detailed computations are possible only in special cases.  One is typically restricted to situations where the perturbation theory is either a solution to a differential equation, or a relatively simple quantum mechanical system \cite{PhysRev.184.1231, bender_anharmonic_1973, Lipatov:1976ny, brezin_perturbation_1977, balian_discrepancies_1978, bogomolny_calculation_1980, zinn-justin_multi-instanton_1981, Balitsky:1985in, jentschura_instantons_2004, Dunne:2013ada, Dunne:2014bca}. Applications of resurgence to observables in  quantum field theories are typically only possible if additional symmetries are present that make the problem tractable \cite{bogomolny_large_1977,Gukov:2016njj, Dunne:2016jsr, Dunne:2013ada,Dunne:2014bca, Dunne:2016nmc}. 

In this paper, we investigate the perturbative expansion of observables in a genuine quantum field theory: the Liouville conformal field theory in two dimensions.  Liouville CFT has been studied extensively in the literature due to its application in string theory \cite{POLYAKOV1981207}, connection to 4D gauge theories \cite{Alday:2009aq}, relations with intersection theory and Seiberg-Witten theory \cite{Matone:1993tj, Matone:1995rx, Bertoldi:2004cc}, and role in describing universal behavior of 2D conformal field theories and semiclassical 3D gravity \cite{Collier:2019weq,Cardy:2017qhl,Chandra:2022bqq,Collier:2023fwi,Chandra:2024vhm}. Understanding the non-perturbative structure of Liouville is therefore of great interest. Moreover, it is a genuine (although simple) theory of gravity, since it may be interpreted as a quantum theory of negatively curved surfaces in two dimensions. One advantage is that Liouville theory comes with a continuously tunable parameter $b$ which is related to the central charge $c$ of the CFT, as $c=1+6Q^2$ with $Q=b+1/b$. The limit of large central charge $c$ (or small $b$) is the semi-classical limit, where observables are computed by the action of an appropriate surface of constant negative curvature.  The parameter $b$ (or equivalently $1/c$) plays the role of $\hbar$ in the theory. We can then study the perturbative series in $1/c$, which can be interpreted as the usual Feynman loop expansion. 

Employing a combination of analytical and numerical methods, we show that the perturbative expansion for observables are asymptotic, and the expansion allows one to extract more-or-less complete information about the non-perturbative effects.  In particular, we will identify the complete set of non-perturbative effects which are required in order to render the asymptotic series sensible and produce a finite result.

We will focus on the three-point functions of Liouville theory. In a generic CFT, conformal symmetry fixes the functional form of the three-point functions of primary operators up to a constant factor:
\begin{equation}
    \label{3ptgencft}
    \langle V_{\Delta_{1}}(x_{1}) V_{\Delta_{2}}(x_{2}) V_{\Delta_{3}}(x_{3}) \rangle = \frac{C(\Delta_{1},\Delta_{2},\Delta_{3};c)}{|x_{12}|^{\Delta_{1}+\Delta_{2}-\Delta_{3}}|x_{23}|^{\Delta_{2}+\Delta_{3}-\Delta_{1}}|x_{13}|^{\Delta_{1}+\Delta_{3}-\Delta_{2}}} \,.
\end{equation}
The $C(\Delta_{1},\Delta_{2},\Delta_{3};c)$ are the operator product expansion (OPE) coefficients, or structure constants, which depend on the details of the theory. For a CFT with central charge $c$, these are functions of the choice of operators, which we have labelled here by their scaling dimensions $\Delta_i$ (in Liouville CFT the only nontrivial primary operators are scalars, and we have assumed this in writing the above expression; in a generic CFT there are spinning primaries and the CFT data depends on the spins as well). The OPE coefficients, along with the scaling dimensions $\Delta_i$ and spins of the primary operators, are the data that completely determines the theory. In a generic CFT, the $C(\Delta_{1},\Delta_{2},\Delta_{3};c)$ will be non-trivial function of the $\{\Delta_i,c\}$, and are not known; this is the famous (and difficult) problem of the classification of CFTs. Remarkably, in Liouville theory the three-point functions are known exactly and are given by the DOZZ formula $C_{\rm DOZZ}(\Delta_{1},\Delta_{2},\Delta_{3};c)$ \cite{Dorn:1994xn,Zamolodchikov:1995aa} (see equation (\ref{DOZZ})). We will be interested in the semi-classical expansion of the DOZZ formula, which will take the schematic form
\begin{equation}
    C_{\rm DOZZ}(\Delta_{1},\Delta_{2},\Delta_{3};c) \approx e^{-c S_{\rm cl}(\Delta_{1},\Delta_{2},\Delta_{3})}\sum_{n=0}^{\infty} \frac{1}{c^{n}} \, f^{(n)}(\Delta_{1},\Delta_{2},\Delta_{3})\, .
\end{equation} 
Here $S_{\rm cl}(\Delta_{1},\Delta_{2},\Delta_{3})$ is the classical action of a saddle point, and $f^{(n)}(\Delta_{1},\Delta_{2},\Delta_{3})$ are the perturbative corrections. We will compute these perturbative corrections exactly at all orders.

Using this perturbative expansion, we will show that the perturbative coefficients grow factorially, leading to the non-convergence of the series expansion.  We will identify the infinite family of non-perturbative effects which are required to render the series finite; these are the instantons of Liouville theory.  This will be accomplished using the method of Borel resummation, where the location of singularities in the Borel plane is related to an appropriate instanton action. 
These instantons are complex saddle points of the path integral formulation of Liouville theory, which can be found alternatively by solving the semiclassical Liouville equation with appropriate operator insertions:
\begin{equation}\label{eq:semiclassical Liouville equation}
    \partial_{a}\partial^{a}\varphi(x) = 8\lambda e^{\varphi(x)} - 8\pi\sum_{i=1}^{n}\eta_{i}\delta^{(2)}(x-x_{i}) \,,
\end{equation}
Here the conformal weights of the operators are related to the $\eta_i$ as $\Delta_i \approx \frac{2\eta_i (1-\eta_i)}{b^2}$ for $b\rightarrow0$ and $\lambda=\pi\mu b^2$, where $\mu$ is the cosmological constant in the Liouville action (\ref{Liouv_action1}). Solutions to this equation for the case of the two- and three-point function have been studied previously in the literature for different ranges of the conformal weights $\eta_i$ \cite{Zamolodchikov:1995aa,Hadasz:2003he,Hadasz:2003kp,Harlow:2011ny}. For certain values of the $\eta_i$, the solutions are related to two-dimensional hyperbolic manifolds. 
Crucially, the instanton solutions which we identify are both complex (in the sense that $\varphi$ is a complex function) and multi-valued.  However, they are not new; these solutions were anticipated using other techniques by Harlow, Maltz and Witten (HMW) \cite{Harlow:2011ny}. The fact that these instantons show up explicitly in a resurgence analysis of DOZZ was also anticipated in \cite{Dunne:2015eaa, Behtash:2015loa}. Importantly, we will find that only a subset of the HMW solutions are necessary. Moreover the asymptotic series appears to be Borel summable in the sense that none of the singularities of the Borel transform lie in a vicinity of the real axis.  

Our conclusion is that Liouville CFT is an example of a genuine quantum field theory -- and a theory of quantum gravity, no less -- where the perturbative and non-perturbative effects can be assembled into a finite answer.  Our hope is that this can provide guidance to more sophisticated examples with more complicated degrees of freedom and couplings.

\subsection*{Outline}
The paper is organized as follows. In section \ref{sec:review}, we review the basics of Liouville theory, the DOZZ formula, and classical solutions to Liouville. Section \ref{sec:res_dozz} provides the main details and results of the computation. We study the perturbative expansion of the three-point function in the semiclassical limit and show that the perturbative series is asymptotic and identify the singularities in the Borel plane. 
We also provide some numerical checks, using a Pad{\'e} approximation where the structure in the Borel plane can be investigated systematically.
In section \ref{sec:c0_analytic} we show that one recovers the information about the complex saddles by studying the analytic properties of the DOZZ formula in the complex $c$ plane.  This structure has appeared in other contexts (see e.g. \cite{Dunne:2012ae}), where expansions around different saddle points are related in a trans-series expansion.

\section{Preliminaries}
\label{sec:review}

\subsection{Review of Liouville Theory}
Here, we give a brief review of Liouville theory and establish the notation that will appear in the rest of the paper. For more details, we refer the reader to for example \cite{Teschner:2001rv,Seiberg:1990eb,Nakayama:2004vk}. Liouville theory is a two-dimensional conformal field theory. The action is\footnote{The $\varphi$ that appears in (\ref{eq:semiclassical Liouville equation}) is related to the field $\phi$ below by $\phi = \frac{\varphi}{2b}$.}
\begin{equation}
    \label{Liouv_action1}
    S_L[\phi] = \frac{1}{4\pi}\int_{\Sigma} d^{2}x \, \sqrt{h} \left(h^{ab}\partial_{a}\phi \, \partial_{b}\phi + Q R \phi + 4\pi\mu\, e^{2b\phi} \right).
\end{equation}
where $\phi$ is the Liouville field. The above action is defined on a two-dimensional surface $\Sigma$ with  (background) metric $h_{ab}$; the Ricci scalar $R$ is computed with respect to this metric. The parameter $Q$ is known as the background charge is related to $b$ as $Q= b + 1/b$ and $\mu$ is the Liouville cosmological constant. The field $\phi$ transforms non-trivially under conformal transformation $z\rightarrow z'(z)$ as
\begin{equation}
    \phi(z,\bar{z}) \rightarrow \phi'(z',\bar{z}') = \phi(z,\bar{z}) - \frac{Q}{2} \log \left( \frac{\partial z'}{\partial z} \frac{\partial \bar{z}' }{\partial \bar{z}} \right)
\end{equation}
The normalizable primary operators of Liouville are scalars that may loosely be thought of as built out of the field $\phi$ as  $V_{\alpha}(x) = \exp(2\alpha \phi(x))$, where $\alpha=\frac{Q}{2}+iP$.  Here $P\in\mathbb{R}_{\ge0}$ is a continuous parameter which labels the primary operators.  We refer to $\alpha$ or $P$ interchangeably as Liouville momenta in the following.  The full spectrum is 
\begin{equation}
    \mathcal{H} = \bigoplus_{\alpha} \mathcal{H}_{\alpha} \otimes \bar{\mathcal{H}}_{\alpha}
\end{equation}
where $\mathcal{H}_{\alpha}$ denotes the highest weight Virasoro representation built on the primary $V_\alpha$. Note that the spectrum is continuous (i.e. Liouville theory is non-compact) and that the identity operator does not appear in the spectrum.  So Liouville differs from the typical compact CFT in that it does not have a normalizable vacuum state.  In many contexts it is also convenient to consider non-normalizable operators, where $\alpha$ is real. The conformal dimension of the primary operator $V_\alpha$ is $h = \bar{h} = \alpha(Q-\alpha)$. Liouville theory is a conformal field theory with central charge is 
\begin{equation}
    c = 1 + 6Q^{2}
\end{equation}
The interesting physical quantities in the quantum theory are the correlation function of the primary operators:
\begin{equation}
    \label{corr_ver}
    \langle V_{\alpha_{1}}(x_{1}) \dots V_{\alpha_{n}}(x_{n}) \rangle = \int [D\phi] \, e^{-S_L[\phi]} \prod_{i=1}^{n} V_{\alpha_{i}}(x_{i})
\end{equation}
These can, at least in principle, be computed in an arbitrary background $\Sigma$ and with an arbitrary number of insertions.  In this paper we will focus primarily on the three point functions in flat space.

\subsubsection*{Structure constants (DOZZ)}
In a CFT, the data that determines the theory are the spectrum and the 3-point functions (i.e. OPE coefficients) of the primary operators (\ref{3ptgencft}). All higher point functions, as well as the partition function on an arbitrary surface, are completely determined by these data. Remarkably, in Liouville theory an explicit expression for the OPE coefficients $C_{\rm DOZZ}(\alpha_1,\alpha_2,\alpha_3) := \langle V_{\alpha_{1}}(\infty) V_{\alpha_{2}}(1) V_{\alpha_{3}}(0) \rangle$ is known. This is the famous DOZZ formula:
\cite{Zamolodchikov:1995aa,Dorn:1994xn}.
\begin{equation}
    \label{DOZZ}
    C_{\rm DOZZ}(\alpha_1,\alpha_2,\alpha_3) = \frac{\left(\pi \mu \gamma(b^2)b^{2-2b^2}\right)^{(Q-\sum_i\alpha_i)/b} \, \Upsilon'_b(0)\Upsilon_b(2\alpha_1)\Upsilon_b(2\alpha_2)\Upsilon_b(2\alpha_3)}{\Upsilon_b(\sum_i\alpha_i-Q)\Upsilon_b(\alpha_1+\alpha_2-\alpha_3)\Upsilon_b(\alpha_2+\alpha_3-\alpha_1)\Upsilon_b(\alpha_3+\alpha_1-\alpha_2)}
\end{equation}
The special functions appearing in this expression are
\begin{equation}
    \gamma(x) = \frac{\Gamma(x)}{\Gamma(1-x)} \,,\quad \log \Upsilon_{b}(x) = \int_{0}^{\infty} \frac{dt}{t} \, \left( e^{-t}(Q/2 - x)^2 - \frac{\sinh^{2}( (Q/2 - x)t/2) }{ \sinh(tb/2)\sinh(t/2b) } \right)
\end{equation}
Note that, because the spectrum is continuous, there is some freedom in the choice of the normalization of the operators $V_\alpha$.  In this expression, 
the normalization has been chosen so that the primary operator two-point function is
\begin{equation}
\langle V_{\alpha_{1}}(0) V_{\alpha_{2}}(1) \rangle = 
	C_{\rm DOZZ}(P_1,P_2,\id) = 2\pi\left[\delta(P_1+P_2)+S(P_1)\delta(P_1-P_2)\right],
\end{equation}
where $S(P)$ is the known as the reflection amplitude
\begin{equation}
	S(P) = (\pi \mu \gamma(b^2)b^{2-2b^2})^{Q-2\alpha}\frac{\Upsilon_b(2\alpha)}{\Upsilon_b(2\alpha-Q)}.
\end{equation}
We will find it convenient to work with a slightly different normalization, where we take the three point coefficients to be
\begin{equation}
    \label{C0ope}
    C_{0}(\alpha_{1},\alpha_{2},\alpha_{3}) = \frac{\Gamma_{b}(2Q)}{\sqrt{2}\Gamma_{b}(Q)^{3}} \frac{\prod_{\pm_{1}\pm_{2}\pm_{3}} \Gamma_{b} \left(\frac{Q}{2}\pm_{1}(\alpha_{1}-\frac{Q}{2})\pm_{2}(\alpha_{2}-\frac{Q}{2})\pm_{3} (\alpha_{3}-\frac{Q}{2}) \right) }{ \prod_{i=1,2,3}\Gamma_{b}(2\alpha_{j})\Gamma_{b}(2Q-2\alpha_{j}) }
\end{equation}
Here 
\begin{equation}
    \begin{split}
        \log \Gamma_{b}\left(x\right) &= \int_{0}^{\infty} \frac{dt}{t} \left[ \frac{e^{-xt} - e^{-Qt/2}}{(1-e^{-bt})(1-e^{-t/b})} - \frac{\left( Q/2 - x \right)^{2}}{2}e^{-t} - \frac{(Q/2 - x)}{t}  \right]
    \end{split}
\end{equation}is related to the $\Upsilon_b(x)$ appearing above by
\begin{equation}
    \Upsilon_{b}(x) = \frac{1}{\Gamma_{b}(x)\Gamma_{b}(Q-x)}
\end{equation}
With this choice of normalization the two-point function of primary operators is
\begin{equation}
\langle V_{\alpha_{1}}(0) V_{\alpha_{2}}(1) \rangle = 
	C_0(P_1,P_2,\id) = \rho_0(P_1)^{-1}\left[\delta(P_1-P_2) + \delta(P_1+P_2)\right].
\end{equation}
where $\rho_0(P) = 4\sqrt{2}\sinh(2\pi b P)\sinh(2\pi b^{-1}P)$. 
This expression for $\rho_0(P)$ is the usual density of states associated with Cardy's formula.
The above form of the DOZZ formula has made appearances in recent studies of 2D CFTs, where it describes the averaged OPE coefficients of heavy operators in every conformal field theory
\cite{Collier:2019weq} (see also \cite{Collier:2016cls,Cardy:2017qhl,Collier:2017shs}).
It also appears in 3D gravity, where it is related to the statistical distribution of coupling constants \cite{Chandra:2022bqq} and the action of hyperbolic manifolds \cite{Collier:2023fwi,Collier:2024mgv}. 

Using $C_{\rm DOZZ}$ or $C_0$ one can construct the higher point correlation functions by performing a conformal block decomposition and inserting factors of $C_{\rm DOZZ}$ together with the appropriate Virasoro conformal blocks. For instance, the four-point function is
\begin{equation}
	\langle V_1(0)V_2(z,\bar z)V_3(1)V_4(\infty)\rangle = \frac{1}{2} \int_{\mb{R}} \frac{dP}{2\pi} C_{\rm DOZZ}(P_1,P_2,P) C_{\rm DOZZ}(-P,P_3,P_4)|\mc{F}_{h_P}(z)|^2
\end{equation}
where the Virasoro block $\mc{F}_{h_P}(z)$ is completely fixed by the $\alpha_i$, the central charge $c$, and the cross-ratio
\begin{equation}
    z = \frac{(z_1-z_2)(z_3-z_4)}{(z_1-z_4)(z_3-z_2)}
\end{equation}

\subsubsection*{Semiclassical Limit}
We are interested in the semiclassical limit of the Liouville correlation functions. This is defined by taking the central charge $c\rightarrow\infty$ which corresponds to $b\rightarrow0$.\footnote{One could equivalently take $b\to\infty$ but we will focus on the $b\to0$ limit for definiteness.} 
Our goal is to study the perturbative expansion around the saddle points of the Liouville path integral. 

To have a well defined semiclassical limit, we define the rescaled Liouville field $\varphi = 2b\phi$.  We will take the background metric to be flat: $h_{ab} = \delta_{ab}$. The action (\ref{Liouv_action1}) becomes
\begin{equation}
    S_L[\varphi] = \frac{1}{16\pi b^{2}}\int d^{2}x \, \left(\partial_{a}\varphi\partial^{a}\varphi + 16\lambda e^{\varphi} \right)
\end{equation}
where $\lambda=\pi\mu b^2$. We see that $b^2$ now plays the role of $\hbar$ in the theory.  Thus $b^2$ will be the loop-counting parameter, and the $b\to 0$ limit is the familiar semi-classical limit of any quantum theory.

In terms of the rescaled field, the primary operators are $V_{\alpha} = \exp(\alpha\varphi/b)$. For the insertions of $V_{\alpha}$ to have a non-trivial effect in the path integral (\ref{corr_ver}) at the semi-classical level, we therefore need $\alpha$ to scale as $1/b$.  So we will take
\begin{equation}
    \alpha = \eta/b
\end{equation} where $\eta$ is held fixed as we take $b\to0$.  This ensures that the primary operator insertions will contribute to the classical equations of motion.
The path integral (\ref{corr_ver}) now takes the form
\begin{equation}
    \label{corr_ver_2}
     \int [D\phi] \, \exp(-\frac{1}{16\pi b^{2}}\int d^{2}x \, \left(\partial_{a}\varphi(x)\partial^{a}\varphi(x) + 16\lambda e^{\varphi(x)}  - 16\pi\sum_{i}\eta_i\varphi(x) \delta^{(2)}(x-x_i)\right) )
\end{equation}
The saddle points of this path integral are found by solving the Liouville equation
\begin{equation}
    \label{liou_eom}
    \partial_a \partial^a \varphi = 8\lambda e^{\varphi} - 8\pi\sum_i \eta_i \delta^{(2)}(x-x_i)
\end{equation}
One can then solve this equation away from the operator insertions, imposing appropriate boundary conditions (discussed below) at the $x_i$.

This semi-classical limit makes clear the interpretation of Liouville theory as a two dimensional theory of gravity.
When $\varphi$ solves (\ref{liou_eom}), the metric
\begin{equation}
ds^2 = e^{\varphi}\sum_{a=1,2}(dx^a)^2
\end{equation} has constant negative curvature.  This can be seen be noting that the two dimensional Ricci scalar is $R:=-e^{-\varphi}\partial_a \partial^a \varphi = -8\lambda$.
The operator insertions correspond to conical defects or geodesic hyperbolic boundaries, depending on the value of $\eta$.
Thus Liouville theory can be interpreted as a quantum theory of hyperbolic geometry in two dimensions.

The behavior near the operator insertions requires some care.  The solutions to (\ref{liou_eom}) when plugged back in to (\ref{corr_ver_2}), lead to divergences at the $x_i$. To regulate this, one needs to introduce a cutoff. The proper `renormalized' action  to compute the correlators of primary operators is given by \cite{Zamolodchikov:1995aa, Harlow:2011ny}
\begin{equation}
    \langle V_{\eta_{1}}(x_1) \dots V_{\eta_{n}}(x_n) \rangle = \int [D\varphi] \, e^{-S_{L}[\varphi]}
\end{equation}
\begin{equation}
    \label{liou_action_2}
    \begin{aligned}
        b^{2}S_{L} &=\, \frac{1}{16\pi}\int_{D'} d^{2}x \, \left(\partial_{a}\varphi\partial^{a}\varphi + 16\lambda e^{\varphi} \right) + \frac{1}{2\pi}\oint_{r=R} d\theta \, \varphi \\
        &\hspace{2cm} - \sum_{i=1}^{n}\left( \frac{\eta_{i}}{2\pi}\oint_{d_{i} } d\theta \, \varphi + 2\eta_{i}^{2}\log \epsilon \right)  + 2\log R 
    \end{aligned}
\end{equation}
$D'$ here is a disc of radius $R\, (\rightarrow\infty)$ with small discs $d_{i}$ of radius $\epsilon \,(\rightarrow0)$ cutout at the insertions of vertex operators.

\subsection{Complex Saddles in Liouville: HMW}
We now very briefly review a few features of the solutions to the classical equation of motion (\ref{liou_eom}) of Liouville theory. The action $S_L$ (\ref{liou_action_2}) of these solutions will provide saddle point contributions to the three-point function described above.

Near an operator insertion at the point $x_i$, the delta function dominates the equation of motion (\ref{liou_eom}) and we have
\begin{equation}
    \partial_a \partial^a \varphi(x) \approx - 8\pi \eta_i \delta^{(2)}(x-x_i) \,, \quad (x\rightarrow x_i)
\end{equation}
\begin{equation}
    \varphi(x) \approx -4\eta_i \log(x-x_i)^2 \,, \quad (x\rightarrow x_i)
\end{equation}
The task is then to find a global solution which interpolates between this asymptotic behavior at the operator insertions.  Such solutions can be found analytically for the two- and three-point functions \cite{Zamolodchikov:1995aa}; we refer to this paper and \cite{Harlow:2011ny} for explicit formulas.  
Indeed, the authors of \cite{Harlow:2011ny} showed that there is a {\it unique} real solution provided that we are in the ``physical region" where the $\eta_i$ are real, with $\eta_i<\frac{1}{2}$ and $\sum_i \eta_i > 1$.  When the $\eta_i$ are real but do not satisfy these inequalities this real solution will develop singularities.  In any case -- as expected -- the action $S_0$ of this solution correctly reproduces the behavior of the DOZZ formula in the classical limit:\footnote{We will often write the contribution of saddle points to DOZZ as $e^{-S_0/b^2}$ to make the $b$ dependence explicit. By $S_0$ we mean $b^2 S_{L}$ where the $S_L$ is evaluated at the saddle point.}
\begin{equation}
\langle V_{\eta_{1}}(x_2) V_{\eta_{2}}(x_2) V_{\eta_{3}}(x_2) \rangle \sim e^{-S_0(\eta_1,\eta_2,\eta_3)/b^2}\,, \quad {\rm as}~b\rightarrow0
\end{equation}
The solution described above is real. We are interested in complex solutions as well, following \cite{Harlow:2011ny}.  One type of complex solution is easy to describe: given a real solution $\varphi_{0}$ of (\ref{liou_eom}), one can generate an infinite number of complex solutions by taking
\begin{equation}
    \label{liou_sol_1}
    \varphi_0 \rightarrow \varphi_N = \varphi_0 + 2\pi i N \,, \quad (N\in\mathbb{Z})
\end{equation}
Plugging this into the action $S_{L}$ (\ref{liou_action_2}), we see that shifts of this kind will generate terms of the form $2\pi i N\eta_i/b^2$ (from the $\eta_i \oint_{d_i} \varphi $ term) and $2\pi iN/b^2$ (from the $\oint_R \varphi $ term). We conclude that 
\begin{equation}
    S_L[\varphi_N] - S_L[\varphi_0] = \frac{2\pi iN}{b^2} \left( 1 - \sum_i \eta_i\right)
\end{equation}
The $\frac{1}{b^2}$ prefactor means that this is a shift of the classical part of the action.
This formula captures the main idea: the action of a complex solution differs from that of the real solution only by an integer plus a linear combination of the $\eta_i$'s.

The solutions described above are all single valued.
However, the authors in \cite{Harlow:2011ny} noted the existence of solutions where the Liouville field $\varphi$ is multi-valued. 
These solutions are most easily understood by interpreting Liouville theory as the boundary theory associated with SL$(2,{\mathbb C})$ Chern-Simons (CS) theory \cite{Verlinde:1989ua}.  
A solution of the Liouville equation on $\Sigma$ can be used to construct a flat SL$(2,{\mathbb C})$ connection on a manifold with topology $\Sigma\times I$, where $I$ is an interval.\footnote{Intriguingly, these connections also have an interpretation in terms of 3D gravity: they represent the gravitaitonal path integral of two-boundary Euclidean wormholes, as considered in \cite{Maldacena:2004rf, Chandra:2022bqq,Collier:2023fwi}, which describe the ``disorder averaged" correlation functions appearing in the dual to three dimensional gravity. In \cite{Collier:2023fwi} this was derived by noting that the Liouville correlation function represents the resolution of the identity in the Hilbert space associated with the quantization of the Teichm\"uller component of the moduli space of flat $PSL(2,\mathbb{R})$ connections on $\Sigma$.}  The CS action of such a saddle is equal to the Liouville action, up to a constant.
The operator insertions are interpreted as singularities of the CS background, around which the CS field has non-trivial monodromy.
The multi-valued solutions can then be constructed by performing a large SL$(2,{\mathbb C})$ gauge transformation, i.e. a gauge transformation which is not continuously connected to the trivial one.  
Under such a large gauge transformation the CS action of a flat connection changes, and the result of \cite{Harlow:2011ny} is that the Liouville action will shift by
\begin{equation}
    \label{hmw_saddles}
    S_L \rightarrow S_L + \frac{2\pi i}{b^2} \left( n + \sum_{i} m_i \eta_i \right) 
\end{equation}
Here $n$ and $m_i$ are integers associated with the choice of large gauge transformation, and the $m_i$'s are either all odd or even.  Interpreted in terms of the original Liouville theory, the Liouville field $\varphi$ is now a multi-valued function.

In general, therefore, we expect that all of these complex saddle points could contribute to the DOZZ formula.  Indeed, as was noted by \cite{Harlow:2011ny}, for general complex values of $\eta$ we have 
\begin{equation}
    \langle V_{\eta_{1}}(x_2) V_{\eta_{2}}(x_2) V_{\eta_{3}}(x_2) \rangle \sim \sum_{N\in\mc{I}}e^{-S_N(\eta_1,\eta_2,\eta_3)/b^2}\,, \quad (b\rightarrow0)
\end{equation}
Here $\mc{I}$ denotes a set of labels for the different multi-valued saddle points (i.e. the $n$ and $m_i$). 
The function $S_N(\eta_1,\eta_2,\eta_3)$ (equal to the $b^2 S_L$ evaluated at the classical solution $\phi_N$) is itself a multivalued function of the $\eta_i$, which we will write down below.

Our interest, however, is the perturbative expansion in powers of $b^2$.  Around any given saddle, we can consider not just the leading semi-classical action but also the infinite series of loop corrections, which will be expressed as a power series in $b^2$.
Putting this together, we expect the DOZZ formula in the semi-classical limit to take the general form\footnote{In principle, we could have an overall factor of $(b^2)^{-\alpha}$ in this expansion but as we explain later $\alpha$ will turn out to be zero for DOZZ.} 
\begin{equation}
    \label{dozz_tseries}
    C_{\rm DOZZ}(\eta_1,\eta_2,\eta_3) \sim  \sum_{N\in\mc{I}}\sum_{n=0}^{\infty} e^{-S_N(\eta_1,\eta_2,\eta_3)/b^2}\, A_{n}^{(N)} b^{2n} \,, \quad (b\rightarrow0)
\end{equation}
In general, we expect that the expansion around any given saddle point should be an asymptotic series rather than a convergent one.
Thus any individual term (i.e.~value of $N$) will not make sense on its own in this sum.  Instead, this expression should be interpreted as 
as the trans-series representation of the OPE coefficients in Liouville theory.  The general idea is that the perturbative coefficients $A_{n}^{(N)}$ and the action of the saddle points $S_N$ appearing in (\ref{dozz_tseries}) are not-independent of one another. In particular, the $A_{n}^{(N)}$ for some fixed $N$ at large $n$, encode information about the action $S_{N'}$ and perturbative coefficients $A_{n'}^{(N')}$ for some $N'$ denoting a different saddle. As we will see that later sections, this idea can be made concrete and tested precisely in Liouville theory. 

\section{Resurgence in DOZZ}
\label{sec:res_dozz}

In this section, we will study the semiclassical expansion of the DOZZ formula by considering the limit of large central charge i.e. $c\rightarrow\infty$ or $b\rightarrow0$. We compute the perturbative coefficients of the 3-point correlators in Liouville theory and study their asymptotic behavior. We will then use the techniques of resurgence to understand the nature of the non-perturbative effects contributing to these observables by studying the singularities in the Borel plane.

In appendices, we provide a brief review of resurgence in general. First, in Appendix \ref{sec:reviewres}, we review a few features of Borel resummation which will be required to make sense of equation (\ref{dozz_tseries}). A useful example to consider before we study the full DOZZ formula is to look at the Gamma function $\Gamma(z)$ in the limit of $z\rightarrow\infty$. Many of the features we will encounter in the DOZZ formula can also be observed in the asymptotic expansion of the Gamma function. The details of this can be found in the appendix \ref{appGamma} where we work out the perturbative expansion for $\Gamma(z)$ in the large $z$ limit, compute its Borel transform, and analyze the singularities in the Borel plane making their connections to the non-perturbative effects/saddle points of the $\Gamma(z)$ integral.  A reader wishing to study a simpler example is encouraged to consult appendices  \ref{sec:reviewres} and \ref{appGamma}.

\subsection{Basic Idea}
Before diving into the details of the asymptotic expansion of the DOZZ formula, let us summarize our expectation based on the results of the previous section.
The OPE coefficients given by DOZZ formula in the semiclassical limit should take the form 
\begin{equation}
    C_{\rm DOZZ}(\eta_1,\eta_2,\eta_3) \simeq  e^{-S_{\rm cl}(\eta_1,\eta_2,\eta_3)/b^2}\, \sum_{n=0}^{\infty} A_{n}(\eta_1,\eta_2,\eta_3) b^{2n} \,, \quad (b\rightarrow0)
\end{equation}
where for now we consider only the contribution of the leading saddle, along with perturbative corrections around this saddle denoted $A_n$. We expect the series to be non-convergent with zero radius of convergence due to growth of $A_n$, i.e. 
\begin{equation}
    A_n(\eta_1,\eta_2,\eta_3) \sim n! (S_N(\eta_1,\eta_2,\eta_3))^{-n} \,, \quad (n\rightarrow\infty)
\end{equation}
Here $S_N(\eta_1,\eta_2,\eta_3)$ is the difference between the action of the first subleading instanton and that of the leading saddle. We anticipate (and will see below) that this instanton should be one of the HMW saddles, so should take the form given in equation (\ref{hmw_saddles}).
Since the $S_N(\eta_1,\eta_2,\eta_3)$ are typically complex, and the coefficients $A_n(\eta_1,\eta_2,\eta_3)$ are real, the complex saddles which control their growth must come in complex conjugate pairs.

More precisely, in the presence of complex saddles, the coefficients $A_n$ at large $n$ will have an expansion of the form \cite{Marino:2012zq,Costin:2021bay}
\begin{equation}
    \label{gen_an}
    A_n \sim \frac{\Gamma(n+\alpha)}{2\pi \abs{S_0}^{n+\alpha}}\left(\abs{c_0} 2\cos((n+\alpha)\theta_{S_0} - \theta_0) + \frac{\abs{c_1}\abs{S_0} 2\cos((n+\alpha-1)\theta_{S_0} - \theta_1)}{n+\alpha-1} + \mathcal{O}(\frac{1}{n^2}) \right)
\end{equation}
where
\begin{equation}
    S_0 = \abs{S_0} e^{i\theta_{S_0}} \,, \quad c_n = \abs{c_n} e^{i\theta_n} 
\end{equation}
and the other constants encode information about the physics of the non-perturbative terms and analytic structure of the Borel sum $B(S)$.  Specifically,
\begin{itemize}
    \item $S_0$ is the location of the leading singularity/closest to the origin in the Borel plane.
    \item $\alpha$ is the exponent of the leading singularity and determines the nature of the branch cut in the Borel plane. With the definition
    \begin{equation}
        B(S) = \sum_{n=0}^{\infty} \frac{A_n}{\Gamma(n+1)}S^{n} \,, \quad C_{\rm DOZZ} = e^{-S_{\rm cl}/b^2}\left(\frac{1}{b^2} \int_{0}^{\infty} dS \, e^{-S/b^2} \, B(S) \right)
    \end{equation}
    the parameter $\alpha$ indicates that the singularity at $S=S_0$ is the form $B(S) \sim 1/(S-S_0)^{\alpha}+\dots$. The case of $\alpha=0$ corresponds to a logarithmic singularity.
    \item $c_0, c_1, \dots$ are related to the perturbative expansion around the first subleading saddle:
    \begin{equation}
        \label{cm_exp}
        i(b^2)^{-\alpha} e^{-S_0/b^2} \sum_{m=0}^{\infty} c_{m} b^{2m}
    \end{equation}
    They also determine the coefficients of the Taylor expansion near the singularity at $S=S_0$ in the Borel plane
    \begin{align}
        B(S = S_0 + z) &\sim \frac{(-z)^{-\alpha}}{2\sin(\pi \alpha)} \sum_{m=0}^{\infty} \frac{c_m \, z^m}{\Gamma(m-\alpha+1)}\,, \quad (\text{branch cut}) \\
        B(S = S_0 + z) &\sim -\frac{\log(-z)}{2\pi} \sum_{m=0}^{\infty} \frac{c_m \, z^m}{\Gamma(m+1)}\,, \hspace{1.4cm} (\text{log cut}) 
    \end{align}
\end{itemize}

\subsubsection*{$\alpha$ in DOZZ}
The value of $\alpha$ (not to be confused with the Liouville momentum parameter) is related to the leading $b^2 (\sim \hbar)$ scaling of the DOZZ formula around a given saddle (\ref{cm_exp}). A quick argument to to observe the leading scaling behavior is as follows: assume that the singularity at $S_0$ is of the form $B(S) \sim (S-S_0)^{-\alpha}$, then the Borel transform around this point would be
\begin{equation}
    \label{alp_scal}
    Z(b^2)|_{\rm inst} = \frac{1}{b^2}\int dS  \, e^{-S/b^2} B(S)  \sim \frac{1}{b^2}\int dS  \, \frac{e^{-S/b^2}}{(S-S_0)^{\alpha}} \sim (b^2)^{-\alpha} e^{-S_0/\hbar} 
\end{equation}
We see that the the leading scaling around the instanton saddle is $(b^2)^{-\alpha}$. A different interpretation for the leading $b^2$ or $\hbar$ can be understood by looking at the moduli space/zero modes around a given instanton. When we compute the one-loop determinant around instantons one needs to evaluate integrals of the form $\int \prod dc_i \, e^{-\sum_i \lambda_i c_i^2}$, where $\lambda_i$'s are the eigenvalues of the appropriate quadratic differential operator/Laplacian i.e. $\nabla \psi_n = \lambda_n \psi_n$ and we decompose the field in terms of this $\varphi_{\rm inst}$ and the fluctuations in terms of the $\psi_n$'s. For the zero modes i.e. the $\lambda_{n'}=0$, the integral over $dc_{n'}$'s can be done separately and each of them contributes to a factor of $\hbar^{-1/2}$. So, the total $\hbar$ scaling around the instanton is expected to be $\sim \hbar^{-D/2}$, where $D$ is the dimension of moduli space or the number of zero modes. Using this we have
\begin{equation}
    Z(\hbar)|_{\rm inst} \sim \hbar^{-D/2} e^{-S_{\rm inst}/\hbar} ~.
\end{equation}
In the case of DOZZ formula, if the external operator dimensions scale with the central charge $h\sim \mc{O}(c)$, then the solutions $\phi_{\rm inst}$ have no moduli \cite{Harlow:2011ny}.  This is the familiar statement that the moduli space of a sphere with three punctures is a point, i.e. the SL$(2,\mathbb{C})$ conformal symmetries of the sphere are completely exhausted by moving the locations of the operators to three fiducial points (usually taken to be $0$, $1$, $\infty$).  So we have $D=0$. Using this in equation (\ref{alp_scal}), we see that expected value of $\alpha$ for the DOZZ formula should be zero i.e. the singularities in the Borel plane should be \textit{log branch} cuts. This allows us to predict that the Borel sum $B(S)$ has the following structure in the complex $S$ plane
\begin{equation}
    B(S) \sim \sum_{N\in\mc{I}} R_{N} \log(S_N-S) 
\end{equation}
where $N$ labels the singular points $S_N$ in $B(S)$ which are the actions of (possibly complex) the HMW saddle points of Liouville theory.

\subsubsection*{$c_n$'s in DOZZ}
For the case of DOZZ formula, the perturbative expansions around different saddles are related in a simple manner, in particular they are exactly equal \cite{Harlow:2011ny}.  The expansion around a saddle labelled by $N$ is
\begin{equation}
    Z_{N} = Z_{\rm cl} \, \exp(\frac{2\pi i}{b^2}\left( n + \sum_i m_i \eta_i \right))
\end{equation}
where $Z_{\rm cl}$ is the expansion around the leading real saddle.
One way to derive this is to note that the different complex saddles in the DOZZ formula arise from the poles in the $\Gamma_b(x)$ appearing in $C_{\rm DOZZ}$.  These functions have poles on the negative real axis, just like the usual Gamma function $\Gamma(x)$.  Just as with the usual Gamma function, the asymptotic series around each one of these poles is identical, differing only by a ``classical action" (in the Gamma function, the location of the pole) but have the same perturbative expansion.  It is straightforward to show that this happens for the $\Gamma_b(x)$ as well. The result of this is that 
\begin{equation}
    c_n = {\cal D} A_n
\end{equation}
where ${\cal D}$ is some fixed constant. This means that the large order behavior of $A_n$'s are determined by low lying $A_n$'s! This is another example of resurgence at work.
In our analysis below, we will verify all the above predictions. We will compute the perturbative coefficients in DOZZ formula explicitly in the semiclassical expansion $b\rightarrow0$. From this, we will show the factorial growth, confirm the location of singularities to be at the HMW locations, and show that they are indeed log branch cuts.

\subsection{Asymptotic Expansion of DOZZ}
We now turn to the computation of the perturbative expansion for the DOZZ formula in the semiclassical limit of $b\rightarrow0$ or $c\rightarrow\infty$. We will use the normalization where the 3-point functions are given by $C_0$
\begin{equation}
    C_{0}(\alpha_{1},\alpha_{2},\alpha_{3}) = \frac{\Gamma_{b}(2Q)}{\sqrt{2}\Gamma_{b}(Q)^{3}} \frac{\prod_{\pm_{1}\pm_{2}\pm_{3}} \Gamma_{b} \left(\frac{Q}{2}\pm_{1}(\alpha_{1}-\frac{Q}{2})\pm_{2}(\alpha_{2}-\frac{Q}{2})\pm_{3} (\alpha_{3}-\frac{Q}{2}) \right) }{ \prod_{i=1,2,3}\Gamma_{b}(2\alpha_{j})\Gamma_{b}(2Q-2\alpha_{j}) }
\end{equation}
where the momenta (i.e.~dimensions) of the external operators are labelled by the $\alpha_i$.
To proceed, we will need the small $b$ expansion of the $\Gamma_b$ functions.

First, as mentioned previously, in order for a vertex operator to have a nontrivial effect on the equations of motion in the semi-classical limit we need $\alpha \propto 1/b$ as $b\rightarrow0$. There are two interesting regimes to consider. The first is $\alpha = \eta/b$ where $0 < \eta < 1/2$.
The second is to write $\alpha = \frac{Q}{2} + iP$, where $P = p/b$ with $p\in\mathbb{R}$.  

Both of these have interpretation in the semi-classical limit in terms of two-dimensional geometries with constant negative curvature. Let us consider the manifold described by the `physical' metric $e^{\varphi}dzd\bar{z}$. Then for the case of $\alpha = \eta/b$, the effect of the operator insertion is to create a conical defect i.e. near the point the geometry looks like that of the plane $dr^2 + r^2 d\theta^2$ but with the identification $\theta \sim \theta + 2\pi (1-2\eta)$. On the other hand the choice $\alpha = \frac{Q}{2} + i\frac{p}{b}$ corresponds to manifolds with boundaries. Each of these boundaries is topologically a circle, and with the hyperbolic metric the geodesic length of the boundary is related the parameter $p$ as $l = 4\pi p$ \cite{Teschner:2002vx}. 

We will consider $\alpha$ to be in either one of these forms, so
\begin{itemize}
    \item $\alpha = \eta/b$
            \begin{equation}
                \label{C0_realeta}
                C_{0}(\eta_1, \eta_2, \eta_3) = \frac{\Gamma_b(2Q)}{ \sqrt{2}\Gamma_b(Q)^3} \frac{\prod_{\pm_1 \pm_2 \pm_3}\Gamma_b(\frac{Q}{2} \pm_1 (\frac{\eta_1}{b} - \frac{Q}{2}) \pm_2 (\frac{\eta_2}{b} - \frac{Q}{2}) \pm_3 (\frac{\eta_3}{b} - \frac{Q}{2}))}{ \prod_{k=1}^3\Gamma_b(\frac{2\eta_k}{b})\Gamma_b(2Q-\frac{2\eta_k}{b})}.
            \end{equation}
    \item $\alpha = \frac{Q}{2} + iP$
            \begin{equation}
                \label{C0_realP}
                C_{0}(P_1, P_2, P_3) = \frac{\Gamma_b(2Q)}{ \sqrt{2}\Gamma_b(Q)^3} \frac{\prod_{\pm_1 \pm_2 \pm_3}\Gamma_b(\frac{Q}{2} \pm_1 iP_1 \pm_2 iP_2 \pm_3 iP_3)}{\prod_{k=1}^3\Gamma_b(Q + 2iP_k)\Gamma_b(Q-2iP_k)}.
            \end{equation}
\end{itemize}
We will study the small $b$ expansion for these separately. 

\subsubsection*{Perturbative expansion of DOZZ for $0<\eta_i<1/2$ as $b^2\rightarrow0$}
We first simplify (\ref{C0_realeta}) using the recursion relation for the $\Gamma_b(z)$ functions:
\begin{equation}
    \Gamma_b(z + b) = \sqrt{2\pi}\, \frac{b^{zb-1/2}}{\Gamma(zb)} \Gamma_b(z)
\end{equation}
\begin{equation}
    \begin{split}
        C_{0}(\eta_1,\eta_2,\eta_3) &= \frac{\Gamma_b(2Q)}{\sqrt{2}\Gamma_b(Q)^3} \,  \frac{ \Gamma_b(\frac{\sum_{i}\eta_i-1}{b} - b)\, \Gamma_b(\frac{2-\sum_{i}\eta_i}{b} + 2b) \, \left(\prod_{i,j,k}' \, \Gamma_b(\frac{\eta_i+\eta_j-\eta_k}{b}) \Gamma_b(\frac{1-\eta_i-\eta_j+\eta_k}{b} + b) \right) }{\left(\prod_{i} \Gamma_b(\frac{2\eta_i}{b})\, \Gamma_b(\frac{2(1-\eta_i)}{b} + 2b) 
        \right)}\\
        \\
        &= \frac{b^{-3/2}}{4\pi^{3/2}} \,  \frac{\prod_{i}\Gamma(2-2\eta_i)}{\Gamma(2-\sum_i \eta_i) \prod_{i,j,k}' \, \Gamma(1-\eta_i-\eta_j+\eta_k) } \frac{\Gamma(\sum_i \eta_ i - 1 - b^2)\,\prod_{i} \Gamma(2-2\eta_i+b^2)} {\Gamma(2+b^2)\Gamma(2-\sum_i\eta_i+b^2)}\, \\
        & \hspace{2cm} \times \frac{\Gamma_b(\frac{2}{b})}{\Gamma_b(\frac{1}{b})^3} \frac{ \Gamma_b(\frac{\sum_{i}\eta_i-1}{b})\, \Gamma_b(\frac{2-\sum_{i}\eta_i}{b} ) \, \left(\prod_{i,j,k}' \, \Gamma_b(\frac{\eta_i+\eta_j-\eta_k}{b}) \Gamma_b(\frac{1-\eta_i-\eta_j+\eta_k}{b}) \right) }{\left(\prod_{i} \Gamma_b(\frac{2\eta_i}{b})\, \Gamma_b(\frac{2(1-\eta_i)}{b}) 
        \right)} \\
    \end{split}
\end{equation}
where $\prod'$ in the first equation indicates cyclic product over $(i,j,k)\in\{(1,2,3),(2,3,1),(3,1,2)\}$. We now study the perturbative expansion of the above expression, which will take the form
\begin{equation}
    C_0(\eta) = (b^2)^{\beta(\eta)} e^{-S_{\text{cl}}(\eta)/b^2} Z_{1-\text{loop}}(\eta)\left( \sum_{n=0}^{\infty}  b^{2n} a_n(\eta)  \right)
\end{equation}
where $S_{\text{cl}}(\eta)$ is the action of the leading or the real saddle point and $Z_{\text{1-loop}}(\eta)$ is the one-loop contribution around the saddle point. These quantities depend on the different $\eta_i$'s, which for brevity we collectively denote $\eta$. The value of $\beta(\eta)$ determines the leading $b^2$ (or $\hbar$) scaling of the perturbative expansion. For the DOZZ formula this vanishes i.e. $\beta(\eta)=0$. This means that the DOZZ formula has a perturbative expansion in integer powers of $b^2$. The $a_n(\eta)$ denote the coefficients of this expansion. The explicit expressions for the action of the real saddle and its one-loop part are
\begin{equation}
    \begin{split}
        S_{\text{cl}}(\eta) &= F\left(\sum_i \eta_i - 1 \right) + \left(F(\eta_1 + \eta_2 - \eta_3)  + {\rm permutations }\right) - \sum_{i} F(2\eta_i) \\
        &\hspace{2cm} - \sum_i (1-2\eta_i)[\log(1-2\eta_i) - 1] - F(0) - 1 \\
    \end{split}
\end{equation}
\begin{equation}
    \begin{split}
        Z_{1-\text{loop}}(p) &=  \frac{1}{4\pi^{3/2}}\left[\frac{\left(\prod_{i}\Gamma(2-2\eta_i)\right)^{3} \left(\Gamma(\sum_i \eta_i - 1)\right)^3  \prod_{i,j,k}' \, \Gamma(\eta_i+\eta_j-\eta_k)}{ \left(\Gamma(2-\sum_i \eta_i)\right)^3 \prod_{i} \Gamma(2\eta_i) \prod_{i,j,k}' \, \Gamma(1-\eta_i-\eta_j+\eta_k) } \right]^{1/2} \\
    \end{split}
\end{equation}
The function $F(z)$ is defined as an integral of $\log\Gamma(z)$.
\begin{equation}
    F(z) = \int_{1/2}^{z} dt\, \log \gamma(t)
\end{equation}
Going back to the geometrical interpretation of the Liouville solutions, the Liouville field that corresponds to the classical action $S_{\text{cl}}$ describes a hyperbolic geometry with $3$ conical defects. The $1$-loop piece is then computed by the inverse square root of the determinant of $(\Delta_0 + \frac{1}{2})$, where $\Delta_0$ is the laplacian on the above hyperbolic manifold. To compute these expressions, we have used the following small $b$ expansions of the $\Gamma_b(x/b)$ \cite{Collier:2018exn}
\begin{align}
    \label{logGammab}
    \log \Gamma_{b}\left(\frac{x}{b} \right) = \frac{\left(\frac{1/2-x}{b} + \frac{b}{2} \right)^2}{2}\log b + \frac{2x-1}{4b^2}\log(2\pi) - \frac{1}{b^2}\int_{1/2}^{x}dt \log \Gamma(t) + C + \sum_{n=0}^{\infty} (b^{2})^{2n+1}  I_{n}
\end{align}
which holds when $\Re(x/b)>0$.
This expansion will also be useful to compute the perturbative expansion of the DOZZ formula. In the above, $C = \frac{1}{2}\log(\frac{\Gamma(x)}{\sqrt{2\pi}})$ and the coefficients $I_n$ are given by
\begin{align}
    \label{In}
    I_{n}(x) &= -\frac{B_{2n+2}}{(2n+2)!} \left( \psi^{(2n)}(x) + ( 1 - 2^{-(2n+1)}) \psi^{(2n)}(1/2) \right) \,, \quad (n\ge0) \\
    &= \frac{B_{2n+2}}{(2n+2)(2n+1)} \left( \zeta(2n+1,x) + ( 1 - 2^{-(2n+1)}) \zeta(2n+1,1/2) \right) \,, \quad (n\ge1) \\
\end{align}
where $\psi^{(m)}(z)$ denotes the polygamma function, defined as the $(m+1)$th derivative of the log gamma function $\frac{d^{m+1}\log\Gamma(z)}{dz^{m+1}}$, $B_{2n}$ are the Bernoulli numbers, and $\zeta(n,x)$ is the Hurwitz zeta function:
\begin{equation}
    \quad B_{2n} = \frac{(-1)^{n+1}2(2n)!}{(2\pi)^{2n}} \zeta(2n) \,,\quad \zeta(n,x) = \sum_{m=0}^{\infty} \frac{1}{(m+x)^{n}}
\end{equation}
where $\zeta(n) = \zeta(n,1)$. 
The result (\ref{logGammab}) can be derived using the integral representation of $\Gamma_b(z)$. We refer the reader to the appendix \ref{app:gammab} for the details. 

Even before delving into details, 
we can already observe the factorial growth of the perturbative expansion. In particular,
\begin{equation}
    \label{asympIn}
    \begin{split}
            I_n &\sim \frac{(-1)^{n}(2n)!}{(2\pi)^{2n+2}} \left( \frac{1}{x^{2n+1}} + 2^{2n+1} \right) \,, \quad (n \gg 1) \\
            &\sim 
            \begin{cases}
                (2n)!(2\pi i x)^{-2n}\,(4\pi^2 x)^{-1} \,,\hspace{0.2cm} x<1/2 \\
                (2n)!(i\pi)^{-2n}\, (2\pi^2)^{-1} \,,\hspace{0.85cm} x>1/2
            \end{cases}  
    \end{split}
\end{equation}
From the above, we expect the non-perturbative effect that should appear in the trans-series to be of the order $\exp(-2\pi ix/b^2)$ or $\exp(-i\pi /b^2)$. As we will see later, once we replace $x$ by the appropriate $\eta$ combinations the former effects are precisely the complex saddle points of the Liouville path integral that the computes the three-point functions of the vertex operators described by HMW \cite{Harlow:2011ny}.

Getting back to the asymptotic expansion of $C_0(\eta)$, by taking an exponential of the above formulas for the $\log\Gamma_b(z)$, we get
\begin{equation}
    \label{c0_eta}
    \begin{split}
        C_0(\eta) &= \exp(\log C_0(\eta)) \\
        &= \exp(-S_{\text{cl}}(\eta)/b^2) \, Z_{\text{1-loop}}(\eta) \, \exp( \sum_{n=0}^{\infty} \left[ (b^{2})^{2n+1}  \tilde{I}_{n}(\eta) + (b^2)^n K_{n}(\eta) \right]  ) \\
        &= \exp(-S_{\text{cl}}(\eta)/b^2) \, Z_{\text{1-loop}}(\eta) \,  \sum_{n=0}^{\infty} b^{2n}\,  a_{n}(\eta)  \\
    \end{split}
\end{equation}
It is easy to check that the $\log b$ term in the exponential cancels when we put together all the $\Gamma_b(z)$ terms along with the $b^{-3/2}$ term in $C_0(\eta)$. From this it follows that $\beta(\eta)=0$. The coefficients $\tilde{I}_n(\eta)$ are linear combination of the $I_{n}(\eta)$ above and the $K_{n}(\eta)$ are the terms from the expansion of the $\log \Gamma(z+b^2)$:
\begin{equation}
    \label{In_tilde}
    \begin{split}
        \tilde{I}_n(\eta) &= 
        \sum_{\epsilon_i = \pm}I_n\left(\frac{1}{2} + \epsilon_1(\eta_1-\frac{1}{2}) + \epsilon_2(\eta_2-\frac{1}{2}) + \epsilon_3(\eta_3-\frac{1}{2})\right) \\
        &\hspace{2cm} - \sum_i \left[I_n(2\eta_i) +  I_n(2-2\eta_i) \right] + I_n(2) - 3I_n(1) \\
    \end{split}
\end{equation}
\begin{equation}
    K_n(\eta)  =  \frac{1}{n!}\left( \psi^{(n-1)}(\sum_i\eta_i-1) + \sum_i \psi^{(n-1)}(2-2\eta_i) - \psi^{(n-1)}(2-\sum_i\eta_i) - \psi^{(n-1)}(2) \right)\, .
\end{equation}
Note that $\log C_0(\eta)$ is the connected piece of the DOZZ formula.  So these are explicit expressions for the $n-$loop contribution to the connected three point function.  Of course, it would be interesting to understand how these can be derived directly from the Feynman loop integrals.\footnote{
One interesting feature of these expressions is that the $n^{th}$ term in this expansion of the connected correlator has degree of transcendentality equal to $n$, coming from the Hurwitz zeta function.  This is much like in the loop expansion of other quantum field theories, such as ${\cal N}=4$ Yang-Mills.}

The $a_n(\eta)$ can be further expressed as a combination of $\tilde{I}_n(\eta)$'s and $K_n(\eta)$'s, but they do not take a particuarly simple form. Nevertheless, it is still easy to see that the $a_n(\eta)$ grow as $n!$ at large $n$ since the dominant contribution at large $n$ comes from the term with the coefficient $\tilde{I}_{\lfloor (n-1)/2 \rfloor}$. The expected exponential piece $S_0^{-n}$ along with the factorial then determines the size of the non-perturbative effects. $S_0$ will now be a function of $\eta$. 

Let us now understand what we expect to find for the subleading saddles using (\ref{In}) and (\ref{In_tilde}).  We have
\begin{equation}
    \begin{split}
        \tilde{I}_n = &(2n)!\frac{2(-1)^{n}\zeta(2n+2)}{(2\pi)^{2n+2}} \sum_{m=0}^{\infty}  \bigg[\sum_{\epsilon_i=\pm} \frac{1}{\left(\frac{1}{2} + \epsilon_1(\eta_1-\frac{1}{2}) + \epsilon_2(\eta_2-\frac{1}{2}) + \epsilon_3(\eta_3-\frac{1}{2}) + m \right)^{2n+1}} \\
        &- \sum_i \left( \frac{1}{(2\eta_i+m)^{2n+1}} + \frac{1}{(2-2\eta_i+m)^{2n+1}} \right) + \frac{1}{(2+m)^{2n+1}} - \frac{3}{(1+m)^{2n+1}} \bigg] + \dots
    \end{split}
\end{equation}
where, $\dots$ indicate the contribution from $\psi^{(m)}$; these subleading terms will not contribute to subleading saddles (i.e. new singularities in the Borel plane) since they do not have factorial growth. Note that only the $x$-dependent term in (\ref{In}) is relevant for the linear combination (\ref{In_tilde}), since the other $x$-independent term cancels once we combine all the $I_n$. These coefficients have the form
\begin{equation}
    \tilde{I}_n \sim (2n)! ( R_0 \abs{S_{0}}^{-(2n+1)} + R_1 \abs{S_{1}}^{-(2n+1)} + \dots) \,, \quad (n \gg 1)
\end{equation}
where $S_{0}, S_{1}, \dots$ are the exponential terms that determine the magnitude of the action for the different saddles. In the above, these actions are all of the form
\begin{equation}
    \begin{split}
        &2\pi N(\sum_i \eta_i -1 + m) \,, 2\pi N(\eta_i + \eta_j - \eta_k + m)\,, 2\pi N(1 + \eta_i - \eta_j - \eta_k + m) \,, 2\pi N(2-\sum_i \eta_i+ m) \,, \\
        &\hspace{2cm} N(2\eta_{i}+ m) \,, N(2-2\eta_{i}+ m)  \,, N(2+m) \,, N(1+m) \,, \quad (N>0, m\ge0)
    \end{split}
\end{equation}
where the factor of $N$ comes from the $\zeta(2n+2)$ multiplying the sum over $m$. Depending on the choice of $\eta_i$, different saddles may be the leading subdominant contribution, so the leading asymptotics of the coefficients $\tilde{I}_n$ will change as a function of $\eta$. For simplicity, let us focus on the case where the $\eta_i$ are equal.  In this case the different saddles that appear have (dropping the $2\pi$ prefactor) actions proportional to
\begin{equation}\label{frodo}
    \begin{split}
        &N(3\eta - 1 + m) \,, N(\eta+ m) \,, N(1-\eta+ m) \,,N(2-3\eta+ m) \\
        & N(2\eta+ m) \,, N(2-2\eta+ m)  \,, N(2+m) \,, N(1+m) \,, \quad (N>0, m\ge0)
    \end{split}
\end{equation}
These are expected contributions coming from the HMW saddles, but crucially they are not {\it all} of the HMW saddles.  We have therefore identified precisely which instanton corrections appear as subleading non-perturbative corrections to the DOZZ formula.
Note that these saddles interchange dominance as $\eta$ is varied.

\begin{figure}[ht]
    \centering
    \includegraphics[scale=.66]{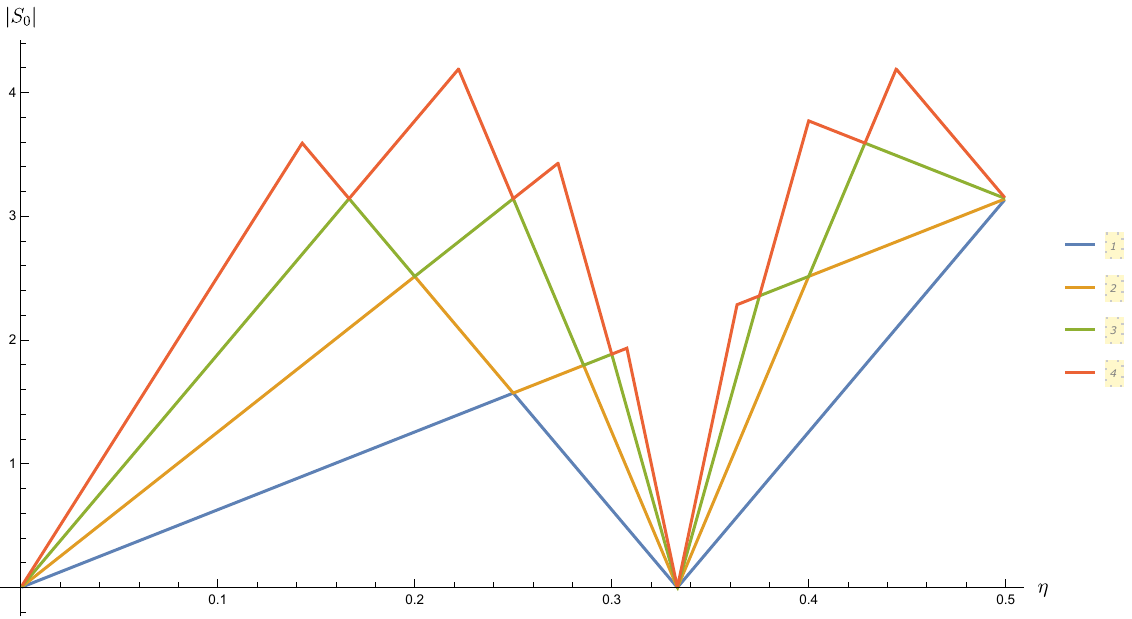}
    \caption{The plot shows the first $4$ subleading saddles for the case of $\eta_i=\eta$. The colors blue, orange, green, and red denote the 1st to 4th dominant saddles respectively. Many of the saddles are degenerate in the sense that there are multiple saddles with the same action; here we plot the first four distinct values of the saddle-point action.}
    \label{fig:sub_pred_same_eta}
\end{figure}

In figure \ref{fig:sub_pred_same_eta}, we plot the actions of the first $4$ subleading saddles as a function of $\eta$, with $0 < \eta < 1/2$. In the later section, when we compute the Borel sum $B(S)$ for the asymptotic perturbative expansion of the DOZZ formula, we verify explicitly that the singular points are located at precisely these locations on the plot.

\begin{figure}[ht]
    \centering
    \includegraphics[scale=0.66]{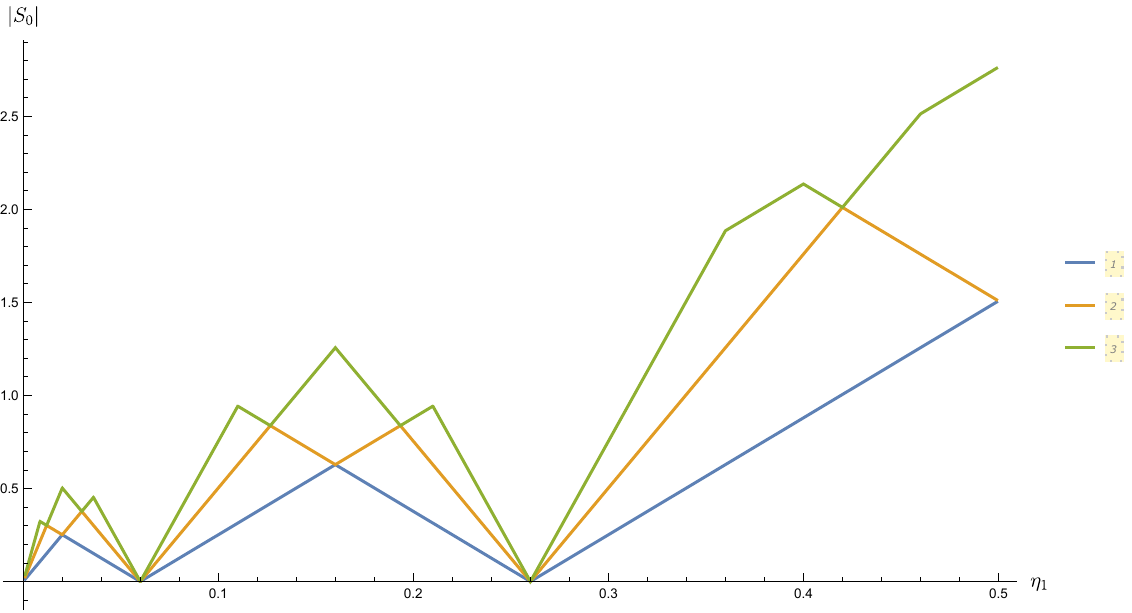}
    \caption{The plot shows the first $3$ subleading saddles for the case of $\eta_2=0.4$ and $\eta_3=0.34$. The colors blue, orange, and green, denote the 1st to 3rd dominant saddles respectively. We see that always 3 different saddles with different actions and no degeneracy.}
    \label{fig:sub_pred_diff_eta}
\end{figure}

As one can imagine the case of different $\eta_i$'s is more interesting and intricate in terms of which saddles exchange dominance. In particular, for generically different values $\eta_i$, the degeneracy of different saddles as in figure \ref{fig:sub_pred_same_eta} is also lifted. In the figure \ref{fig:sub_pred_diff_eta}, we show the behavior of the first $3$ subleading saddles plotted as function of $\eta_1$ for fixed $\eta_2=0.4$ and $\eta_3=0.34$.
There are multiple saddles that exchange dominance in the figure. In particular, the different actions that give rise to the different curves are
\begin{equation}
    \begin{split}
        2\pi \abs{N(\textstyle \sum_i \eta_i - 1)} \,, 2\pi \abs{N(\eta_1 - \eta_2 + \eta_3 - 1)} \,, 2\pi N(2\eta_1) \,, \quad (N=1,2,3) 
    \end{split}
\end{equation}

\subsubsection*{Perturbative expansion of DOZZ for $p_i \in \mb{R}$ as $b^2\rightarrow0$}
We now turn to the case of complex $\alpha$.  In this case the leading semi-classical saddle is a hyperbolic geometry with boundaries rather than conical defects. As in the previous case, we can simplify equation (\ref{C0_realP}) for the structure constant using the recursion relation for the $\Gamma_b(z)$ functions:
\begin{equation}
    \begin{split}
        C_0(p) &= \frac{1}{\sqrt{2}} \frac{\Gamma_b(2Q)}{\Gamma_b(Q)^3} \frac{ \abs{\Gamma_b( (1/2 + i\sum_i p_i)/b + b/2)}^2 \prod_{i,j,k}'\abs{\Gamma_b( (1/2 + i(p_i+p_j-p_k))/b + b/2)}^2 }{ \prod_i \abs{\Gamma_b( (1 + 2ip_i)/b + b)}^2} \\
        &=  \left(\frac{b^{b^2 - \frac{3}{2}}}{16\pi^{7/2}} \prod_{k=1}^{3}\frac{2\pi p_k}{\sinh(2\pi p_k)} \right)  \, \\
        &\quad \quad \times \frac{\Gamma_b(2/b)}{\Gamma_b(1/b)^3}  \frac{ \abs{\Gamma_b( (1/2 + i\sum_i p_i)/b + b/2)}^2 \prod_{i,j,k}'\abs{\Gamma_b( (1/2 + i(p_i+p_j-p_k))/b + b/2)}^2 }{ \prod_i \abs{\Gamma_b( (1 + 2ip_i)/b)}^2} \\
    \end{split}
\end{equation}
We now study the perturbative expansion of the $b$-dependent piece of $C_0(p)$. In the limit $b^2 \rightarrow 0$, this takes the form
\begin{equation}
    C_0(p) = (b^2)^{\beta(p)} e^{-S_{\text{cl}}(p)/b^2} Z_{1-\text{loop}}(p)\left( \sum_{n\ge0}  b^{2n} \tilde{a}_n(p)  \right)
\end{equation}
where $S_{\text{cl}}(p)$ is the action of the leading or the real saddle point and $Z_{\text{1-loop}}(p)$ is the one-loop contribution around the saddle point. We will see that $\beta(\eta)=0$, again implying a perturbative expansion in integer powers of $b^2$. The $\tilde{a}_n(p)$ denote the coefficients of this expansion. The explicit expressions for the action of the real saddle and its one-loop part are
\begin{equation}
    \begin{split}
        S_{\text{cl}}(p) &= F\left( \frac{1}{2} - i\sum_i p_i \right) - \sum_{i} F(1+2ip_i) + \left(F\left( \frac{1}{2} + i(p_1 + p_2 - p_3) \right) + {\rm permutations} \right) \\
        &\hspace{2cm} - \sum_{i}(2ip_k)[\log(2ip_k) - 1] - F(0) - 1  \\
    \end{split}
\end{equation}
\begin{equation}
    \begin{split}
        Z_{1-\text{loop}}(p) &= \sqrt{\pi} \prod_{k=1}^{3} \sqrt{\frac{2\pi p_k}{\sinh(2\pi p_k)}}
    \end{split}
\end{equation}
Going back to the geometrical interpretation of the Liouville solutions, the Liouville field that corresponds to the classical action $S_{\text{cl}}$ describes a hyperbolic geometry with $3$ boundaries whose length is related to $p$ as $l=4\pi p$. The $1$-loop piece is then computed by the inverse square root of the determinant of $(\Delta_0 + \frac{1}{2})$, where $\Delta_0$ is the laplacian on the above hyperbolic manifold.

To compute these expressions we now need the small $b$ expansions of the $\Gamma_b(x/b + b/2)$ in addition to $\Gamma_b(x/b)$ \cite{Collier:2018exn}. This is given by 
\begin{align}
    \log \Gamma_{b}\left(\frac{x}{b} + \frac{b}{2}\right) = \frac{\left(\frac{1}{2}-x \right)^2}{2b^2} \log b + \frac{2x-1}{4b^2}\log(2\pi) - \frac{1}{b^2}\int_{1/2}^{x}dt \log \Gamma(t) + \sum_{n=0}^{\infty} (b^2)^{2n+1}  J_{n}
\end{align}
In the above, the coefficients $J_n(x)$ are given by
\begin{align}
    \label{Jn}
    J_{n}(x) = \frac{B_{2n+2}}{(2n+2)!} \left(1 - \frac{1}{2^{2n+1}} \right) \left( \psi^{(2n)}(x) - \psi^{(2n)}(1/2) \right) 
\end{align}
Just as in the previous case, the $J_n$ grow factorially with $n$ due to the Bernoulli numbers $B_{2n}$. To get the asymptotic expansion of $C_0(p)$ we exponentiate the $\log \Gamma_b$ expansions
\begin{equation}
    \begin{split}
        C_0(p) &= \exp(-S_{\text{cl}}(p)/b^2) \, Z_{\text{1-loop}}(p) \, \exp( \sum_{n=0}^{\infty} (b^{2})^{2n+1}  \tilde{J}_{n}(p)  ) \\
        &= \exp(-S_{\text{cl}}(p)/b^2) \, Z_{\text{1-loop}}(p)  \,  \sum_{n=0}^{\infty} b^{2n}\,  \tilde{a}_{n}(p)  \\
    \end{split}
\end{equation}
Once again, adding up the $\log b$ terms, we find they vanish so that $\beta(p)=0$. The coefficients $\tilde{J}_n$ are
\begin{equation}
    \label{Jn_tilde}
    \begin{split}
        \tilde{J}_n(p) = & \sum_{\epsilon_i = \pm} J_n\left(\frac{1}{2} + i(\epsilon_1 p_1 + \epsilon_2 p_2 + \epsilon_3 p_3) \right) - \sum_{k=1}^{3}\sum_{\epsilon = \pm} I_{n}(1 + \epsilon\, 2ip_k) + I_{n}(2) - 3I_{n}(1)
    \end{split}
\end{equation}
Again, this is a completely explicit formula for the $n$-loop Feynman diagram contribution to the connected three point function.
Similar to our discussion for the case of real $\eta$, we can study (\ref{Jn_tilde}), to understand the behavior of the subleading saddles:
\begin{equation}
    \begin{split}
        \tilde{J}_n = &(2n)!\frac{2(-1)^{n}\zeta(2n+2)}{(2\pi)^{2n+2}} \sum_{m=0}^{\infty} \sum_{\epsilon_i=\pm}\bigg[ -\left(1-\frac{1}{2^{2n+1}}\right)\frac{1}{\left(\frac{1}{2} + i(\epsilon_1 p_1 + \epsilon_2 p_2 + \epsilon_3 p_3) + m \right)^{2n+1}} \\
        &- \left(\sum_{k=1}^{3}\sum_{\epsilon=\pm}  \frac{1}{(1+\epsilon 2ip_k+m)^{2n+1}}\right) + \frac{1}{(2+m)^{2n+1}} - \frac{3}{(1+m)^{2n+1}} \bigg]\, .
    \end{split}
\end{equation}
The possible values of the action for the different saddles are now given as
\begin{equation}
    \label{saddles_realp}
    2\pi N \left(\frac{1}{2} + i\sum_i p_i + m\right)\,, 2\pi N \left(\frac{1}{2} + i( p_i + p_j - p_k) + m \right)\,, 2\pi N \left(1 + 2i p_k + m \right)\,, 2\pi N
\end{equation}
along with their complex conjugates. Note that  the saddles now have actions with both a real and imaginary part. There will again be transitions among these saddles as the $p_i$ are varied.

\subsection{Large \texorpdfstring{$n$}{n} behavior of perturbative coefficients}
We can analyze the coefficients $a_n$ (or $\tilde{a}_n$) analytically and show that they indeed have the form (\ref{gen_an}). Let us consider the case of $0<\eta<1/2$. From (\ref{c0_eta}), we see that for the large $N$, the relevant terms come from $\tilde{I}_{N}$.
\begin{equation}
    a_{N} = \tilde{I}_{\frac{N-1}{2}} + \frac{\tilde{I}_{\frac{N-3}{2}}\tilde{I}_{0}^2}{2} +  \dots \,, (N \text{ - odd}) \,, \quad a_{N} = \tilde{I}_{\frac{N-2}{2}}\tilde{I}_{0} + \tilde{I}_{\frac{N-4}{2}}\tilde{I}_{1} + \dots \,,  (N\text{ - even})
\end{equation}
Furthermore, using the explicit expressions (\ref{In}), (\ref{In_tilde}), the $\tilde{I}_n$ have the form
\begin{equation}
    \tilde{I}_n \sim \frac{2i}{2\pi} \frac{(2n)!}{S_0^{2n+1} }
\end{equation}
where $S_0$ is the exponential piece that determines the location of the first singular point in the Borel plane. Substituting this in the above expressions
\begin{align}
    &a_{N} \sim \frac{2i}{2\pi}\left( \frac{\Gamma(N)}{{S_0^{N}}} + \frac{\Gamma(N-2)}{{S_0^{N-2}}}\frac{\tilde{I}_{0}^2}{2} + \dots \right) \,, \quad (N \text{ - odd}) \\
    &a_{N} \sim \frac{2i}{2\pi}\left(\frac{\Gamma(N-1)}{{S_0^{N-1}}}\tilde{I}_{0} + \frac{\Gamma(N-3)}{{S_0^{N-3}}}\tilde{I}_{1} + \dots \right) \,,\quad  (N\text{ - even})
\end{align}
combining these two expressions, we can write
\begin{equation}
    a_N = \frac{\Gamma(n)}{2\pi \abs{S_0}^{N}} \left[ 2\cos(\frac{(N-1)\pi}{2}) + \tilde{I}_{0} \abs{S_0}\frac{ \cos(\frac{(N-2)\pi}{2})}{N-1} + \frac{\tilde{I}_{0}^2 \abs{S_0}^2}{2}\frac{ \cos(\frac{(N-3)\pi}{2})}{(N-1)(N-2)} + \dots \right]
\end{equation}
comparing this with (\ref{gen_an}), we see that the different parameters are
\begin{align}
    &\alpha=0 \,, \quad \theta_{S_0} = \theta_{n} = \pi/2 \\
    &\abs{c_0} = 1 \,, \quad \abs{c_1} = \tilde{I}_{0} \,, \quad  \abs{c_2} = \frac{\tilde{I}_{0}^2}{2} \,,\quad \dots
\end{align}
It is easy to check that the coefficients $\abs{c_n}$ are precisely the low order loop expansion terms around the original saddle which can be seen by expanding $\exp( \sum_{n=0}^{\infty} (b^{2})^{2n+1}  \tilde{I}_{n}(\eta) )$ and checking the first few terms. A similar analysis for the case of $\tilde{a}_n(p)$ shows they have the form (\ref{gen_an}) with the actions now given by (\ref{saddles_realp}).

We can also numerically evaluate the parameters $c_0, c_1, \dots$ using (\ref{In}),(\ref{In_tilde}),(\ref{Jn}), and (\ref{Jn_tilde}). For the case, when all $\eta_i$ are equal, this is shown in table \ref{tab:c0_values}. Note that the values of $c_0$ are not exactly $1$ but rather are positive integers. This has a simple explanation: because the $\eta_i$'s are equal, there are multiple saddle points with exactly the same action. For example, for $\eta < 1/4$ there are 3 saddles with $\eta_i + \eta_j - \eta_k$ and permutations, giving the factor of 3. Similarly, for $1/4 < \eta < 1/3$, the saddles with $1-\sum_i \eta_i$ have degeneracy of $1$, for $1/3 < \eta < 1/2$, the saddles with $\sum_i \eta_i - 1$ likewise have degeneracy  $1$. Finally for $\eta=1/2$, the saddle with $\sum_i \eta_i-1,\, \eta_i+\eta_j-\eta_k,\, 1+\eta_i-\eta_j-\eta_k,\, 2-\sum_i \eta_i$ and permutations has degeneracy $8$.
\begin{table}[ht]
    \centering
    \begin{tabular}{|c|c|c|c|c|}
        \hline
        $\eta$ & $\abs{c_0}$ & $\abs{c_1}$ &  $\abs{c_1/c_0}$ & $\abs{\tilde{I}_0(\eta)}$  \\
        \hline
        $0.10$ & $3$ & $11.452$ & $3.817$ & $3.817$ \\
        $0.15$ & $3$ & $4.590$ & $1.530$ & $3.817$ \\
        $0.20$ & $3$ & $0.727$ & $0.242$ & $0.242$ \\
        $0.27$ & $1$ & $4.145$ & $4.145$ & $4.145$ \\
        $0.30$ & $1$ & $9.563$ & $9.563$ & $9.563$ \\
        $0.35$ & $1$ & $22.552$ & $22.552$ & $22.552$ \\
        $0.40$ & $1$ & $6.018$ & $6.018$ & $6.018$ \\
        $0.45$ & $1$ & $3.528$ & $3.528$ & $3.527$ \\
        $0.48$ & $1$ & $2.873$ & $2.873$ & $2.873$ \\
        $0.50$ & $8$ & $20.910$ & $2.614$ & $2.613$ \\
        \hline
    \end{tabular}
    \caption{Numerical evaluation of $\abs{c_0}$ and $\abs{c_1}$ for different values of $\eta$. The values of $|c_0|$ precisely correspond to the degeneracies of the first subleading saddle points. Taking these degeneracies into account, the ratio of $\abs{c_1/c_0}$ then also agrees with $\tilde{I}_0(\eta)$ which is the 2-loop term $a_{1}(\eta)$ around the original saddle.}
    \label{tab:c0_values}
\end{table}

\subsection{Numerical Computations}
In this section, we numerically confirm the above predictions of the behavior of the perturbative coefficients, and the exhibit explicitly the subleading complex saddles appearing in the DOZZ formula. In all cases we find exact agreement with the predictions discussed above for the case of $0<\eta_i<1/2$ and $p_i\in\mb{R}$.

\subsubsection{Numerics for \texorpdfstring{$0<\eta_i<1/2$}{eta}}
We consider the case where all the $\eta_i$'s are equal. We have
\begin{equation}
    \begin{split}
        C_{0}(\eta) &= A(\eta) b^{-3/2} \frac{\Gamma(3\eta-1-b^2)\,\Gamma(2-2\eta+b^2)^{3}} {\Gamma(2+b^2)\Gamma(2-3\eta+b^2)}\, \frac{\Gamma_b(\frac{2}{b})}{\Gamma_b(\frac{1}{b})^3} \, \frac{ \Gamma_b(\frac{3\eta-1}{b})\, \Gamma_b(\frac{1-\eta}{b})^{3} \, \Gamma_b(\frac{\eta}{b})^{3} \, \Gamma_b(\frac{2-3\eta}{b})}{\Gamma_b(\frac{2\eta}{b})^{3} \, \Gamma_b(\frac{2(1-\eta)}{b})^{3}} \\
    \end{split}
\end{equation}
where we denote $C_{0}(\eta, \eta, \eta) = C_0(\eta)$ for simplicity, and we have defined the $b$-independent part to be
\begin{equation}
    A(\eta) = \frac{1}{4\pi^{3/2}} \,  \frac{\Gamma(2-2\eta)^3}{\Gamma(2-3\eta) \Gamma(\frac{\eta}{b})^3 } 
\end{equation}
The explicit form of the coefficients $a_n(\eta)$ (\ref{c0_eta}) will involve a complicated combination of the $I_n$'s.  We evaluate these numerically and study the behavior at large $n$. From this we can identify the factorial growth, the action of the sub-leading saddles (both the first as well as the further $($sub$)^n$- leading saddles), and the order of the singularities. To compare with our general discussion on the Borel transform, we define
\begin{equation}
    B(S) = \sum_{n=0}^{\infty} d_n(\eta)\, S^{n} \,, \quad d_n = \frac{a_n(\eta)}{\Gamma(n+1)}\, .
\end{equation}
We expect this to have a finite radius of convergence in the Borel plane defined by the complex $S$ variable. Using this, we can then rewrite $C_0(\eta)$
\begin{equation}
    C_0(\eta) = e^{-S_{\text{cl}}(\eta)/b^2} \, Z_{\text{1-loop}}(\eta) \left( \frac{1}{b^2} \int_{0}^{\infty} dS \, e^{-S/b^2} \, B(S) \right)
\end{equation}
since the coefficients $a_n(\eta)$ grow factorially $(\sim n! \abs{S_0}^{-n})$.  We determine the value of $S_0$ by computing quantities like the ratio $\abs{d_{n+1}/d_n}$ as $n\rightarrow\infty$. Here, $S_0$ denotes the magnitude of the first sub-leading (complex) saddle. In practice, it is better to compute a more refined quantity
\begin{equation}
    c_n^2 = \frac{d_{n+1}d_{n-1} - d_n^2}{d_n d_{n-2} - d_{n-1}^2} \,,\quad (n\ge3)
\end{equation}
at large $n$, which has an asymptotic expansion
\begin{equation}
    c_n = \frac{1}{\abs{S_0}}\left(1 + \frac{\alpha-1}{n}\right) + \mathcal{O}\left(\frac{1}{n^2}\right)
\end{equation}
where we have taken $d_n \sim n^{\alpha-1}\abs{S_{0}}^{-n}\cos((n+\alpha)\theta_{S_0} - \theta_0)$. This is the expected behavior since $a_n \sim \Gamma(n+\alpha) \abs{S_{0}}^{-n}\cos((n+\alpha)\theta_{S_0} - \theta_0)$ from (\ref{gen_an}). We have assumed that $\theta_{S_0}, \theta_{0}$ are generic and non-zero (when $\theta_{S_0}=\theta_{0}=0$, then the $\alpha-1$ is replaced by $\alpha-2$). The HMW saddles generically have complex action, as we will verify by evaluating the coefficients $A_n$. To identify the value of $\abs{S_0}$ and $\alpha$, we plot the $c_n$'s against $1/n$ and fit a straight line to it. 

\begin{figure}[t!]
    \centering
    \includegraphics[scale=0.4]{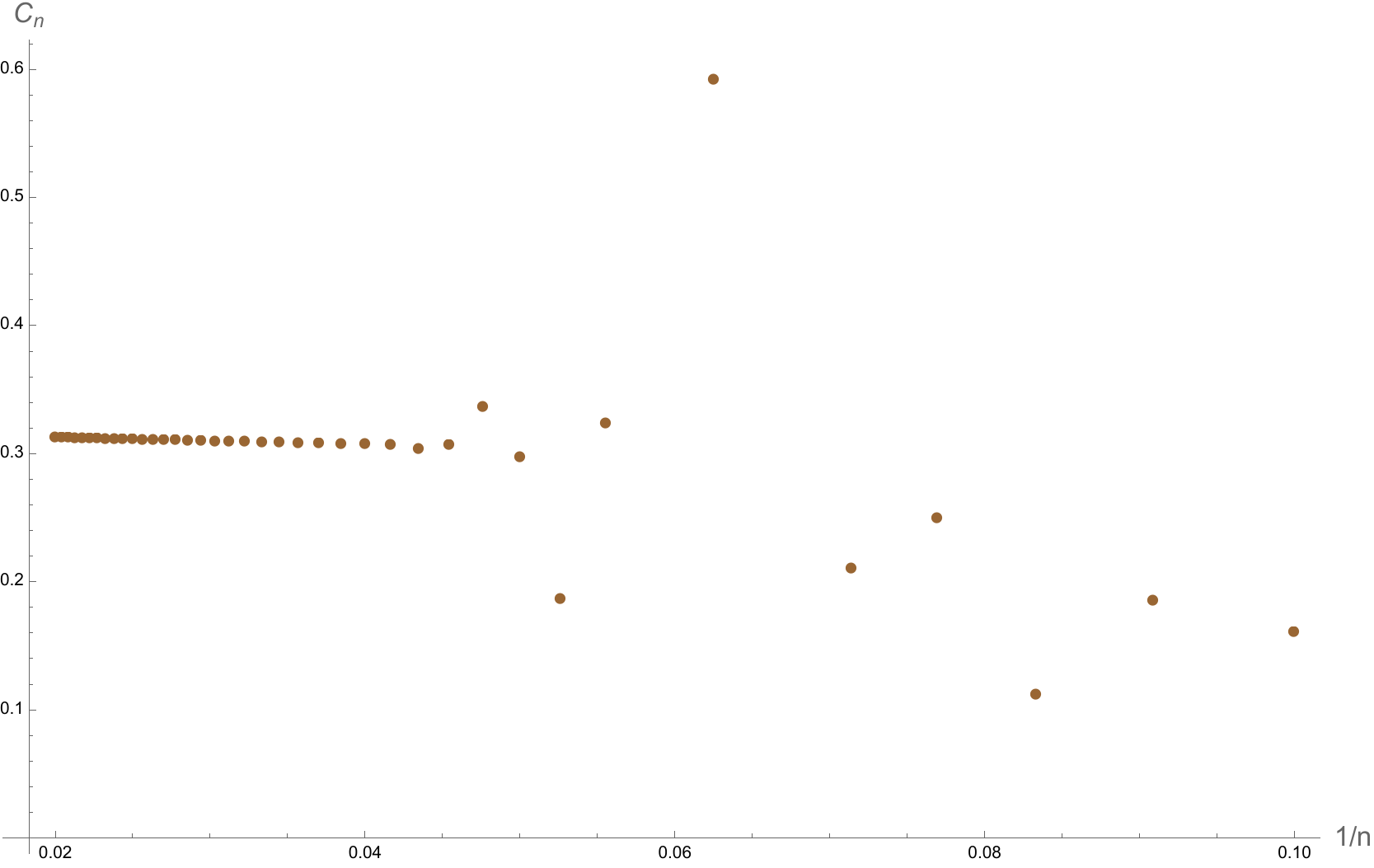}
    
    \vspace{0.5cm}
    \includegraphics[scale=0.4]{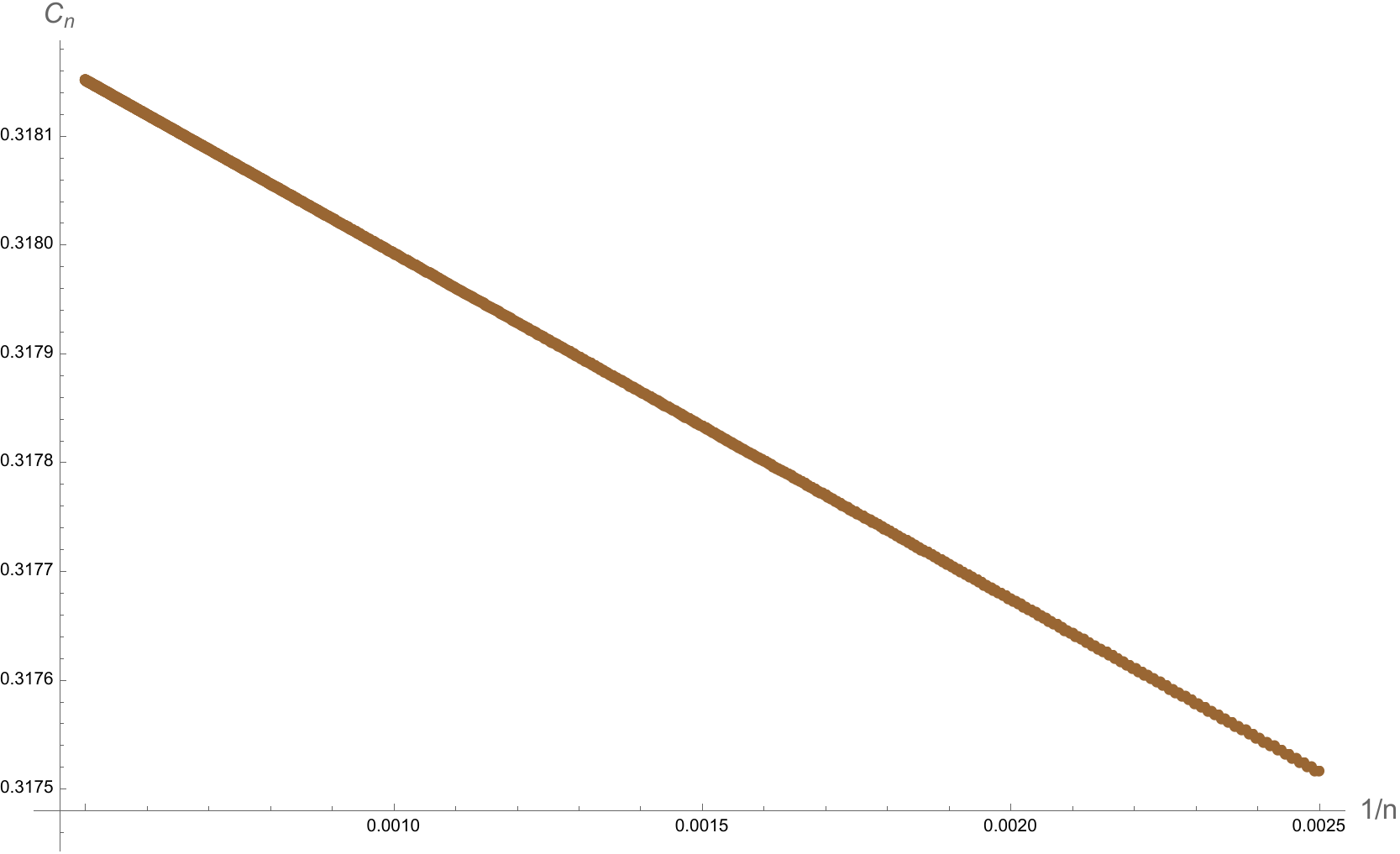}
    \caption{The plot of $c_n$ vs $1/n$ for the Liouville three point function at the threshold $C_0(\eta=1/2)$. In the first plot we show the $c_n$'s from $n=10$ to $50$. As we can see, after a few fluctuations, the $c_n$ settle on a straight line. In the second plot, we zoom into this region and plot terms between $n=400$ to $2000$. The straight line fit $y=0.3183098 - 0.3182315\, x$ can be used to extract the values of $\abs{S_0}$ and $\alpha$. Using this, we get $\abs{S_0} = 3.1415932$ and $\alpha = 0.0002460$.}
    \label{fig:C0_eta0.5}
\end{figure}

As a first check, we consider the case of $\eta=1/2$, shown in the figure \ref{fig:C0_eta0.5}. From the plot, we see that the $c_n$'s for large $n$ do lie on a straight line and the value of the action and the order of the singularity are
\begin{equation}
    \boxed{\abs{S_0} = 3.1415932 \,, \quad \alpha = 0.0002460}
\end{equation}
It is not hard to notice that above value of $S_0$ is approximately $\pi$. This is just $\abs{2\pi i\eta }$ or $\abs{2\pi i(1-3\eta) }$ which is the difference between the action of real saddle $S_{\text{cl}}$ and the complex HMW saddles when $\eta=1/2$. It is also interesting to note that we can also demonstrate that the singularity in the Borel plane is a \textit{log branch cut} since $\alpha$ is close to zero to a high precision.

\begin{figure}[t!]
    \centering
    \includegraphics[scale=0.4, trim = 40 0 0 0, clip]{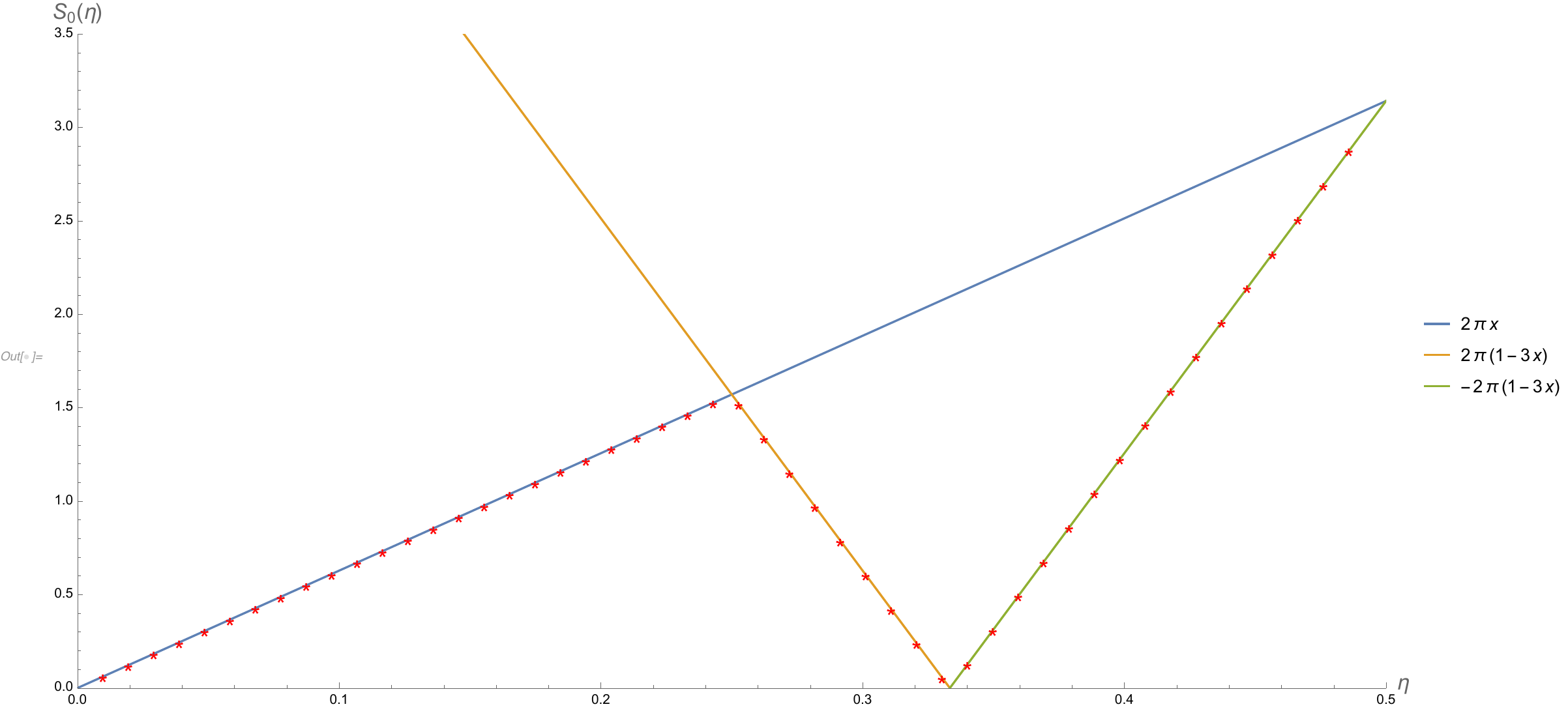}
    \caption{The value of the saddle point action $\abs{S_0(\eta)}$ computed using the asymptotic expansion of the DOZZ formula around the real saddle. The red markers indicate numerical results, and the straight lines denote the actions of HMW saddles \cite{Harlow:2011ny}. These saddles interchange dominance at $\eta=1/4$ and $\eta=1/3$.}
    \label{fig:HMW_s0_poles}
\end{figure}

We can repeat the above analysis for different values of $\eta$ between $0$ and $1/2$. In the table \ref{tab:C0_eta}, we state the results of the numerics for the action of the sub-leading saddles and the order of singularities. Again, we observe in each case that the numerical values agree with the expected value to high precision.
\begin{table}[t!]
    \centering
    \begin{tabular}{|c|c|c|c|c|}
        \hline
        $\eta$ & $\abs{S_0(\eta)}$ & $\alpha$ & $\abs{2\pi i(1-3\eta)}$ & $\abs{2\pi i\eta}$ \\
        \hline
        $0.50$ & $3.1415924$ & $0.0002460$ & $3.1415927$ & - \\
        $0.40$ & $1.2566373$ & $0.0002486$ & $1.2566370$ & - \\
        $0.35$ & $0.3141593$ & $0.0002307$ & $0.3141593$ & - \\
        $0.22$ & $1.3823009$ & $0.0001416$ & - & $1.3823008$ \\
        $0.10$ & $0.6283185$ & $0.0000372$ & - & $0.6283185$ \\
        \hline
    \end{tabular}
    \caption{The magnitude of the action of the sub-leading saddles for different values of $\eta$. In each case, they agree to a high accuracy to the value as determined in \cite{Harlow:2011ny} by solving the Liouville equation of motion in the presence of $3$ conical defects. The analytic form of the sub-leading action changes from $\abs{2\pi i(1-3\eta)}$ to $\abs{2\pi i\eta}$ at $\eta=1/4$ since the lowest action term dominates the asymptotics of the perturbative coefficients and $\eta=1/4$ is the location where the different solutions have the same action.}
    \label{tab:C0_eta}
\end{table}
It is helpful to visualize this results by the plotting $S_0(\eta)$ against the different values of $\eta$. This is shown in figure \ref{fig:HMW_s0_poles} where we have added multiple other points in the range $0<\eta<1/2$ and in each case with find a match between the HMW saddle points and the numerical result. 

\subsection*{Pad\'e-Borel}
In principle, even the information about the sub-sub-leading saddles (and in general $(\text{sub})^n$-leading) is encoded in the coefficients $a_n(\eta)$. However, since the perturbative coefficients are expected to behave as $a_n \sim n!S_0^{-n}f_0 + n!S_1^{-n}f_1 + \dots$, the term with $S_1$ is exponentially suppressed compared to $S_0$ (assuming $\abs{S_0}<\abs{S_1}$). This makes extracting their values somewhat tricky. 

To proceed numerically, we can make progress by using the method of Pad\'e approximants to extract the location of the $(\text{sub})^n$-leading singular points in the Borel plane and the magnitude of the action of these saddles. We will use the Pad\'e approximants to the Borel sum $B(S)$ to find these singular points.

In general, the idea behind the Pad\'e approximants is as follows: given a Taylor expansion for a function $f(x)$, we approximate this function as a ratio of two polynomials of degree $m$ and $n$ respectively. The Pad\'e approximant of order $[m/n]$ is defined as
\begin{equation}
    f(x) = \sum_{n=0}^{N}a_n x^{n} \implies P_{[m/n]}f(x) := \frac{P_m(x)}{Q_n(x)} = \frac{p_0 + p_1 x + \dots + p_m x^m}{1 + q_1 x + \dots + q_n x^n}
\end{equation}
The coefficients $p_k$ and $q_k$ are determined by the requirement that the Taylor expansion of the Pad\'e approximant matches the Taylor expansion of the function $f(x)$ up to order $m+n$. We will use the diagonal Pad\'e approximants i.e. $m=n=\lfloor N/2 \rfloor$.

\begin{figure}[ht]
    \centering
    \includegraphics[scale=0.45]{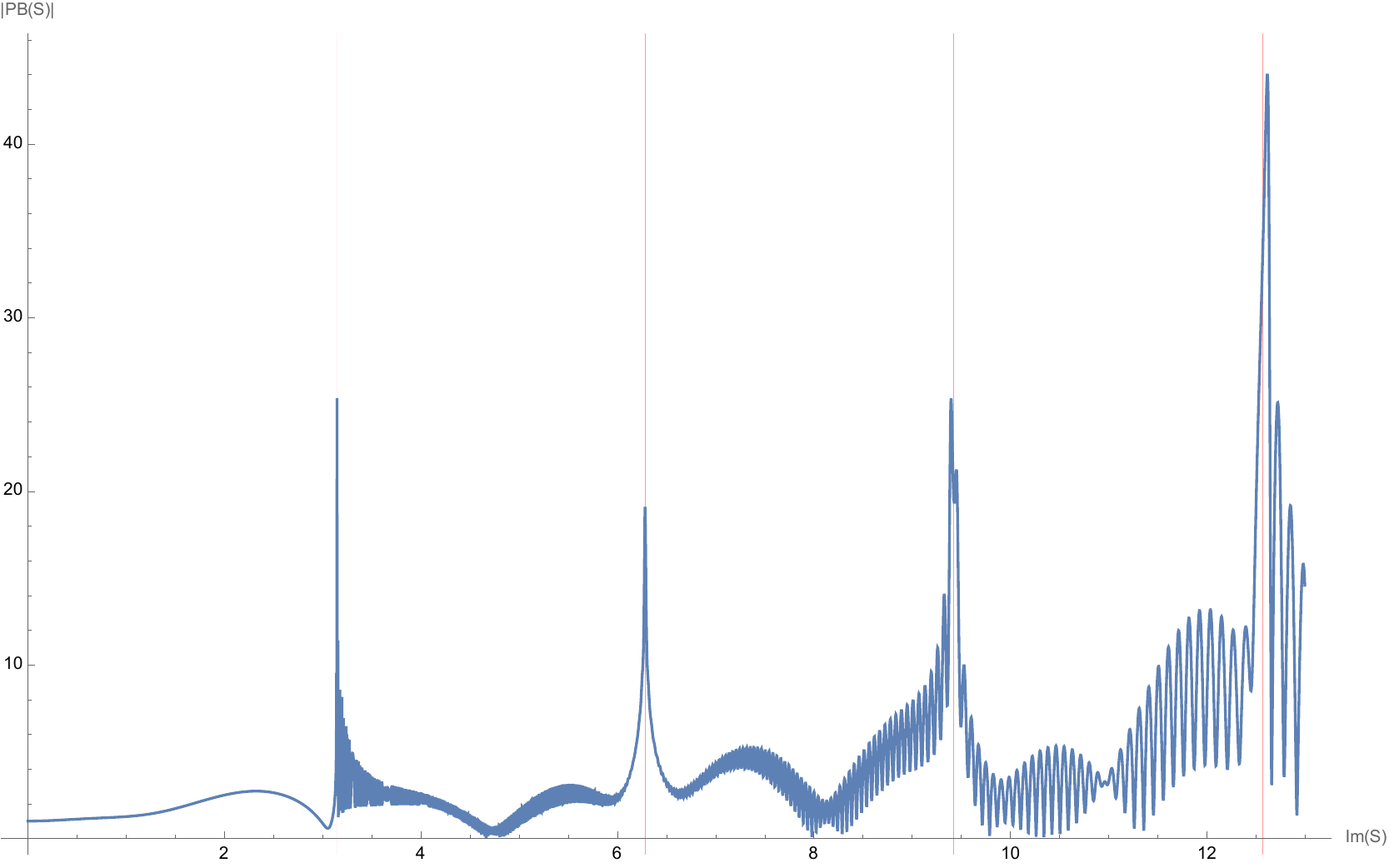}
    \caption{The plot of $\abs{PB(S)}$ for the Pad\'e-Borel sum of the Liouville three point function $C_0(\eta)$ for $\eta=1/2$. Here, we plot $\abs{PB(S)}$ along the imaginary $S$ axis. The red vertical lines mark the integer multiple of $2\pi i(1-3\eta)$ or $2\pi i\eta$. We can see that the function has a singular behavior at precisely the expected actions of complex saddles. For the above figure, we compute $PB(S)$ by using a $\lfloor 1400,1400\rfloor$ Pad\'e approximant.}
    \label{fig:C0_pade_im}
\end{figure}

\begin{figure}[ht]
    \centering
    \includegraphics[scale=0.35]{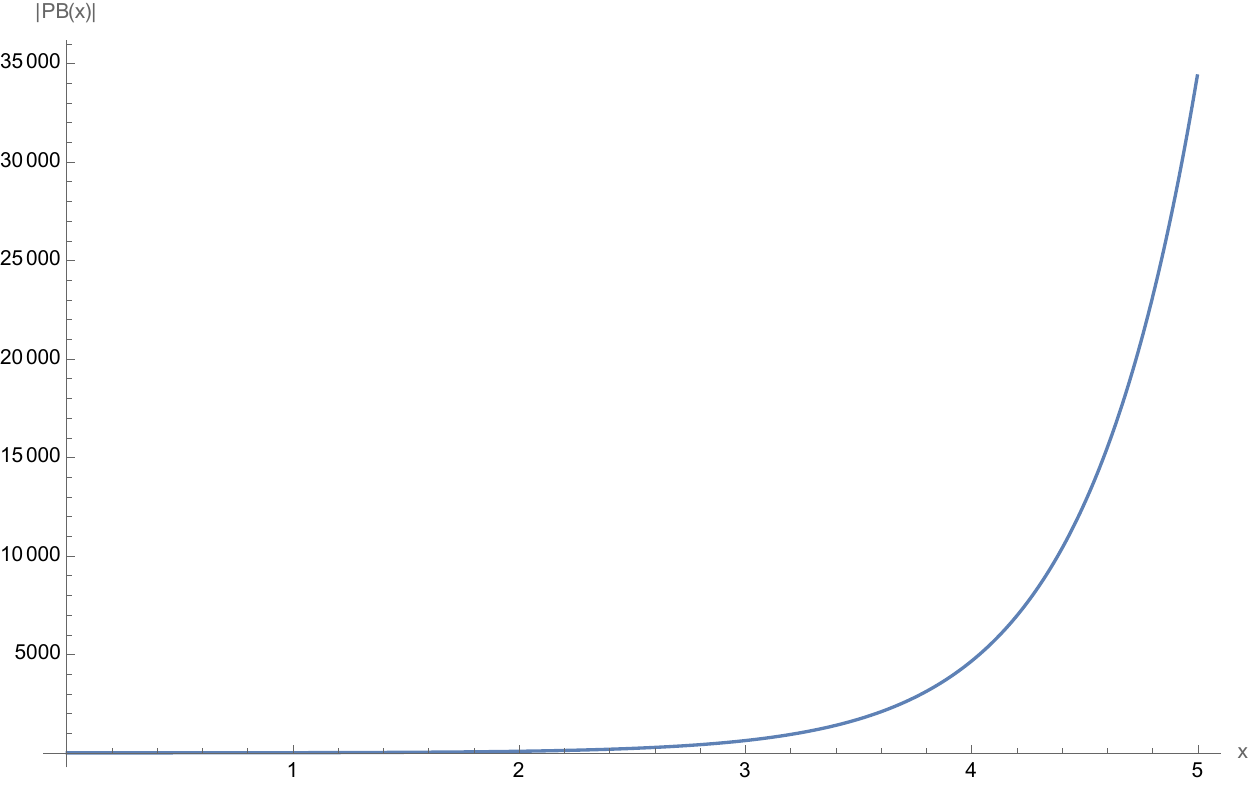}
    \hspace{0.5cm}
    \includegraphics[scale=0.35]{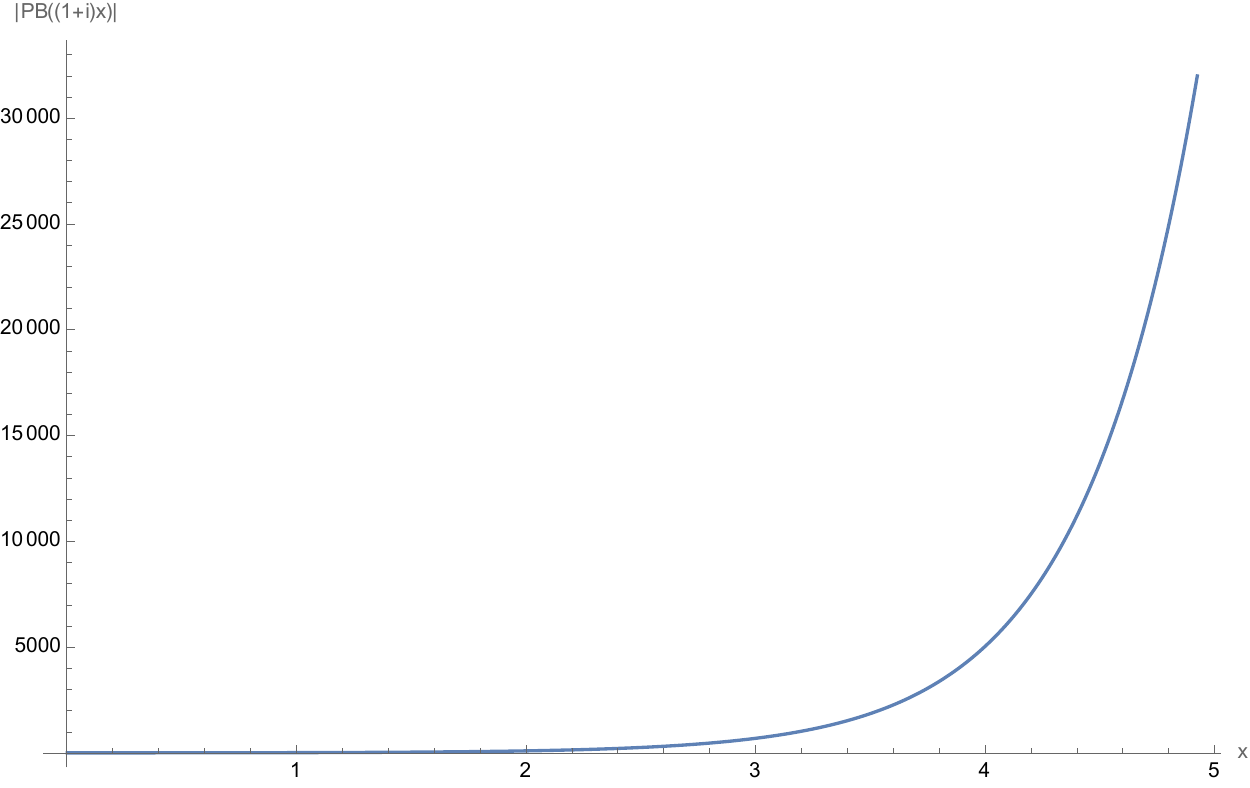}
    \caption{The plot of $\abs{PB(S)}$ for the Pad\'e-Borel sum of $C_0(\eta=1/2)$. Here, we plot $\abs{PB(S)}$ along $S=x$ and $S=(1+i)x$ direction with $x$ being a positive real number. We see that there is no singular behavior along these directions. For the above figure, we compute the $PB(S)$ by using a $\lfloor 1400,1400\rfloor$ Pad\'e approximant.}
    \label{fig:C0_pade_re}
\end{figure}

\begin{figure}[ht]
    \centering
    \includegraphics[scale=0.4, trim = 40 0 0 0, clip]{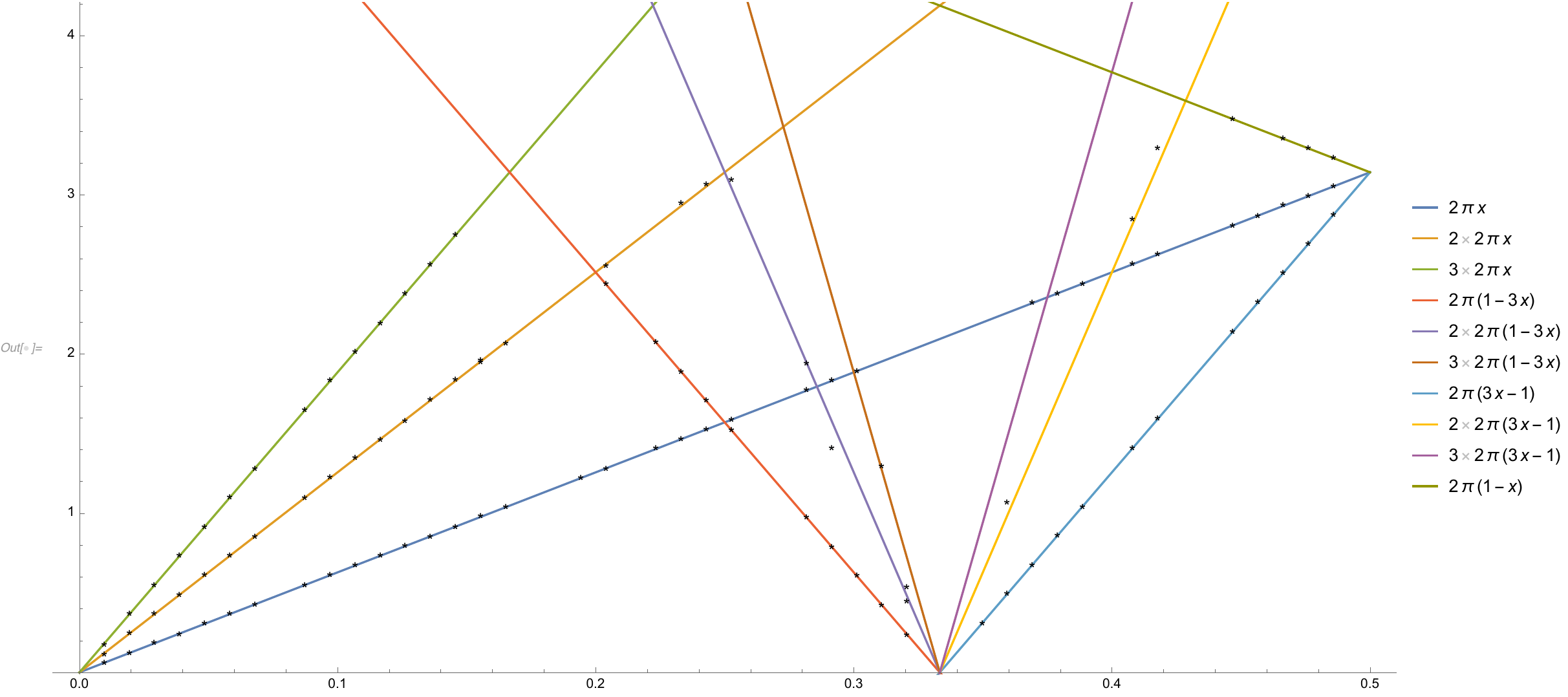}
    \caption{The value of the $(\text{sub})^n$-leading saddle point action computed using the asymptotic expansion of the DOZZ formula around the real saddle. The markers indicate the location of the poles in Pad\'e-Borel approximants and the different straight lines denote the HMW saddles \cite{Harlow:2011ny}. 
    }
    \label{fig:HMW_s0_poles_sub}
\end{figure}

Applying this to the Borel sum of the DOZZ formula $B(S)$, we can rewrite it as a rational function i.e.
\begin{equation}
    B(S)|_{\rm trunc.} = \sum_{n=0}^{N} d_n S^{n} \implies PB(S) = \frac{P_{[N/2]}(S)}{Q_{[N/2]}(S)}
\end{equation}
Using this we can study the analytic behavior of $B(S)$ in the Borel plane. We can plot this function as a function of the complex $S$ variable and study its behavior. In particular, we find that $PB(S)$ has singular points along the imaginary axis and nowhere else. This is shown in figure \ref{fig:C0_pade_im} and figure \ref{fig:C0_pade_re} for $\eta=1/2$ where we plot $\abs{PB(S)}$ along different lines passing through the complex $S$-plane. From the figure, we see that the poles lies precisely at the locations $2\pi i(1-3\eta)$ or $2\pi i\eta$ and its integer multiples. As in the figure \ref{fig:HMW_s0_poles}, we can repeat the analysis for different values of $\eta$ and identifying the $(\text{sub})^n$-leading saddles. This is shown in figure \ref{fig:HMW_s0_poles_sub} \footnote{In the figure what we actually use to search for the location of the singular points is the Pad\'e-approximant for the derivative of the Borel sum i.e.~$B'(S) = \sum_{n=1}^{\infty} d_n S^{n-1}$ rather than $B(S)$ directly. Numerically, this is helpful as the peaks in the Pad\'e-approximant are stronger and helps in better identification. This is expected since previously we saw that the singular point in $B(S)$ are expected to be $\log(S_0 - S)$ singularities and a derivative converts them to poles $\frac{1}{(S_0 - S)}$.}.

The results from these numerics make it clear that the method of computing the perturbative coefficients and approximating the Borel sum using a Pad\'e approximants can be implemented successfully to identify the singular points in the Borel plane. The singular points can further be identified as exactly the action of the non-perturbative effects that can be computed using saddle point methods.

\subsubsection{Numerics for \texorpdfstring{$p\in\mb{R}$}{p}}

\begin{figure}[t!]
    \centering
    \vspace{0.5cm}
    \includegraphics[scale=0.4]{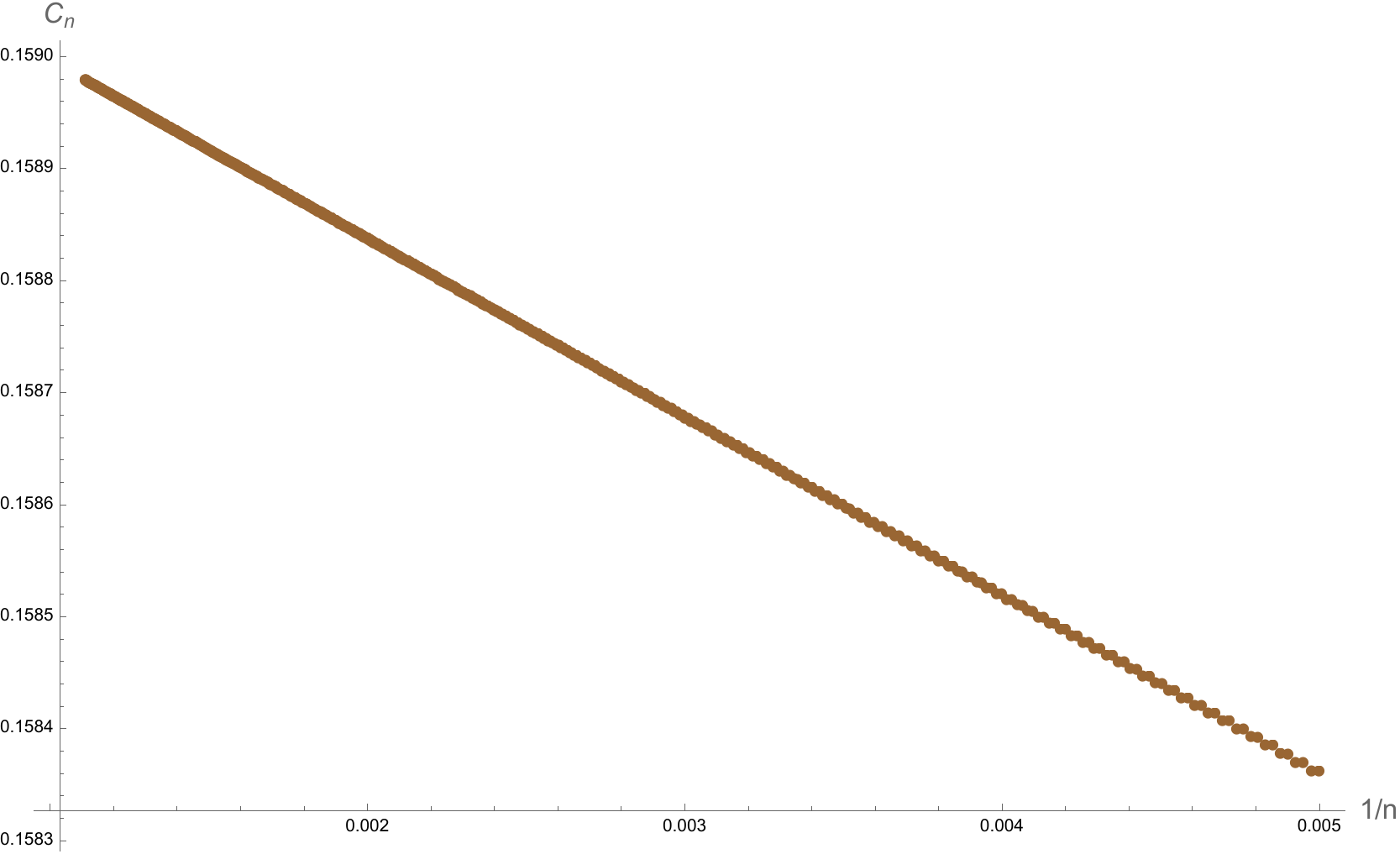}
    \caption{The plot of $c_n$ vs $1/n$ for the Liouville three point function $C_0(p=10\pi)$. In the plot we show the terms between $n=200$ to $900$. The straight line fit $y=0.1591549 - 0.1591430 x$ can be used to extract the values of $\abs{S_0}$ and $\alpha$. Using this, we get $\abs{S_0} = 6.2831861$ and $\alpha = 0.0000748$. This exactly matches the action of an HMW saddle.}
    \label{fig:C0_p_pi}
\end{figure}

As in the above example, we start with the case where all the $p_i$'s are equal. The $\tilde{a}_n(p)$ are then linear combinations of the $\tilde{J}_n$'s which grow factorially. We are interested in extracting the exponential piece $S_{0}^{-n}$ by defining the Borel transform and computing it numerically using the techniques used above. Defining
\begin{equation}
    \tilde{B}(S) = \sum_{n=0}^{\infty} \tilde{d}_n(p)\, S^{n} \,, \quad \tilde{d}_n = \frac{\tilde{a}_n(p)}{\Gamma(n+1)}
\end{equation}
the full $C_0(p)$ can then written as
\begin{equation}
    C_0(p) = e^{-S_{\text{cl}}(p)/b^2} \, Z_{\text{1-loop}}(p) \left( \frac{1}{b^2} \int_{0}^{\infty} dS \, e^{-S/b^2} \, \tilde{B}(S) \right)
\end{equation}
We construct the $c_n$'s as before and plot them as a function of $1/n$ to find $S_0$. In figure \ref{fig:C0_p_pi}, we show this for the case of $p=\pi$. From the plot, we see that the $c_n$'s for large $n$ do lie on a straight line and the value of the action and the order of the singularity are
\begin{equation}
    \boxed{\abs{S_0} = 6.2831861 \,, \quad \alpha = 0.0000748}
\end{equation}
From this, we note that the magnitude of the action is $2\pi$ again with a logarithmic branch cut in the Borel plane. In particular, for most values of $p$, the leading action is always equal to $2\pi$. However, as we will see below, using the method of Pad\'e-Borel, there are sub-leading saddles that do indeed depend on $p$ and unlike the case of real $\eta$, we will see that the saddles do not all lie on the imaginary axis and can have a nontrivial
real part. 
 This is exactly as expected from (\ref{saddles_realp}). 

\subsection*{Pad\'e-Borel}
\begin{figure}[t!]
    \centering
    \includegraphics[scale=0.27]{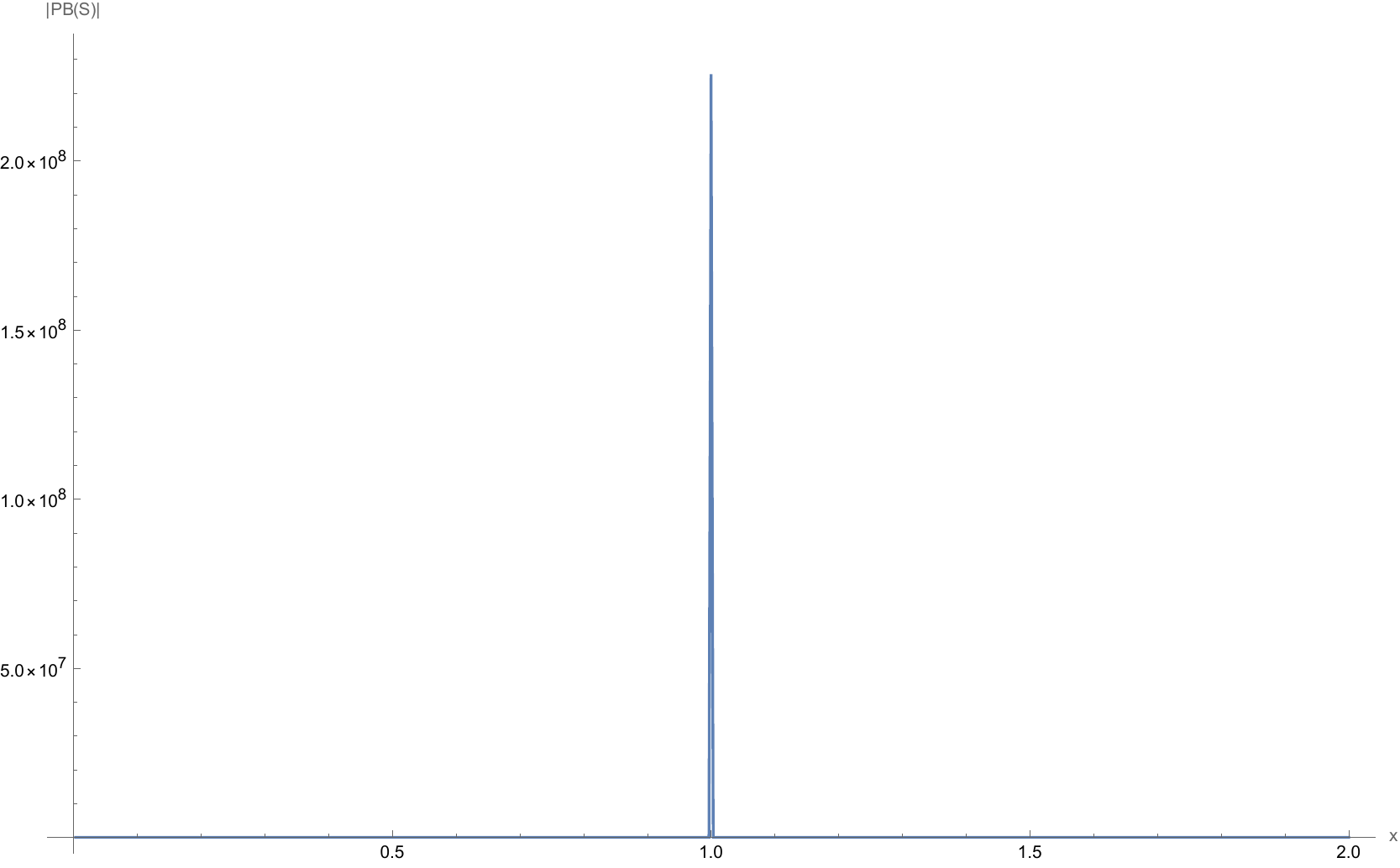}

    \vspace{0.3cm}
    \includegraphics[scale=0.27]{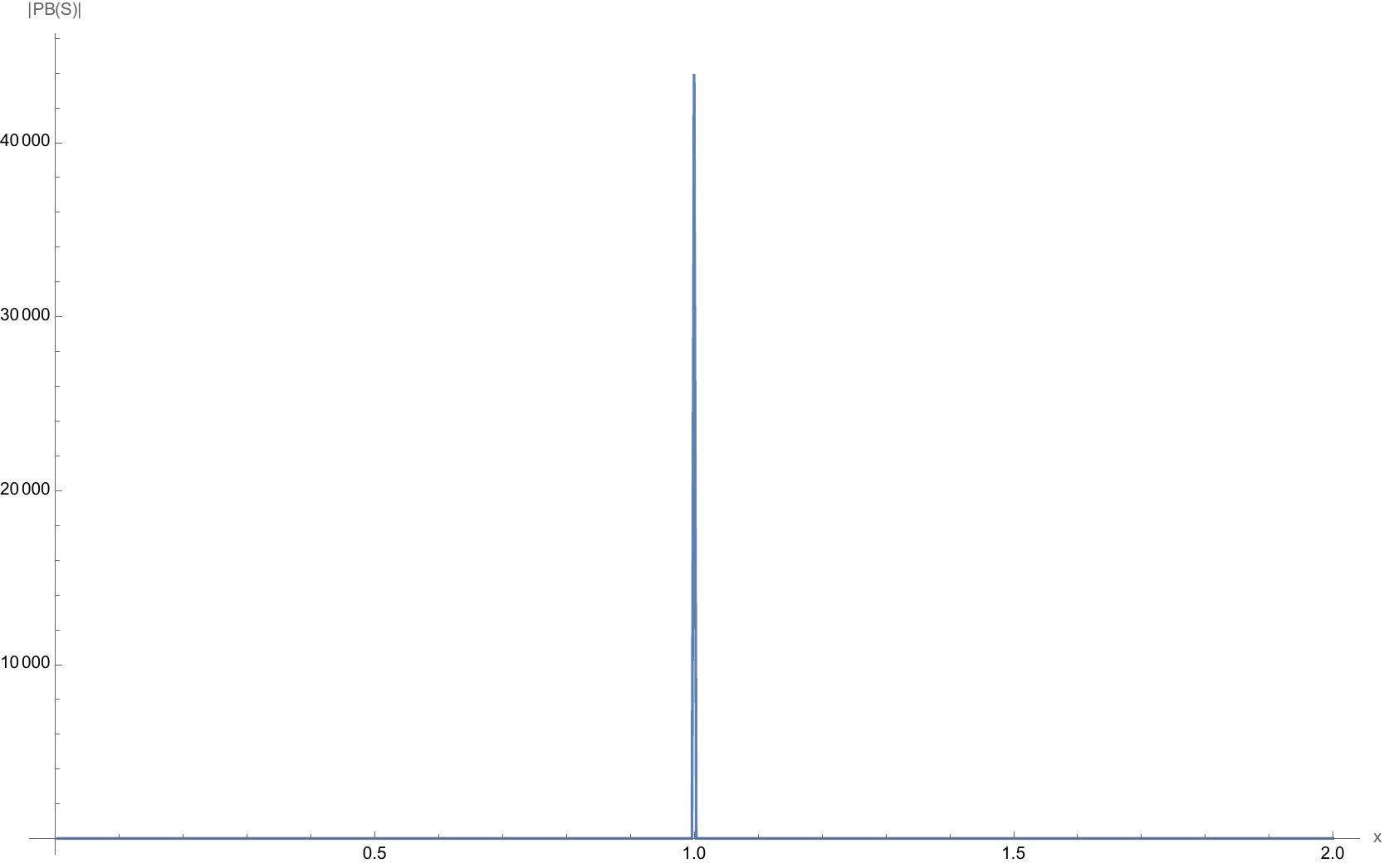}
    \hspace{0.5cm}
    \includegraphics[scale=0.27]{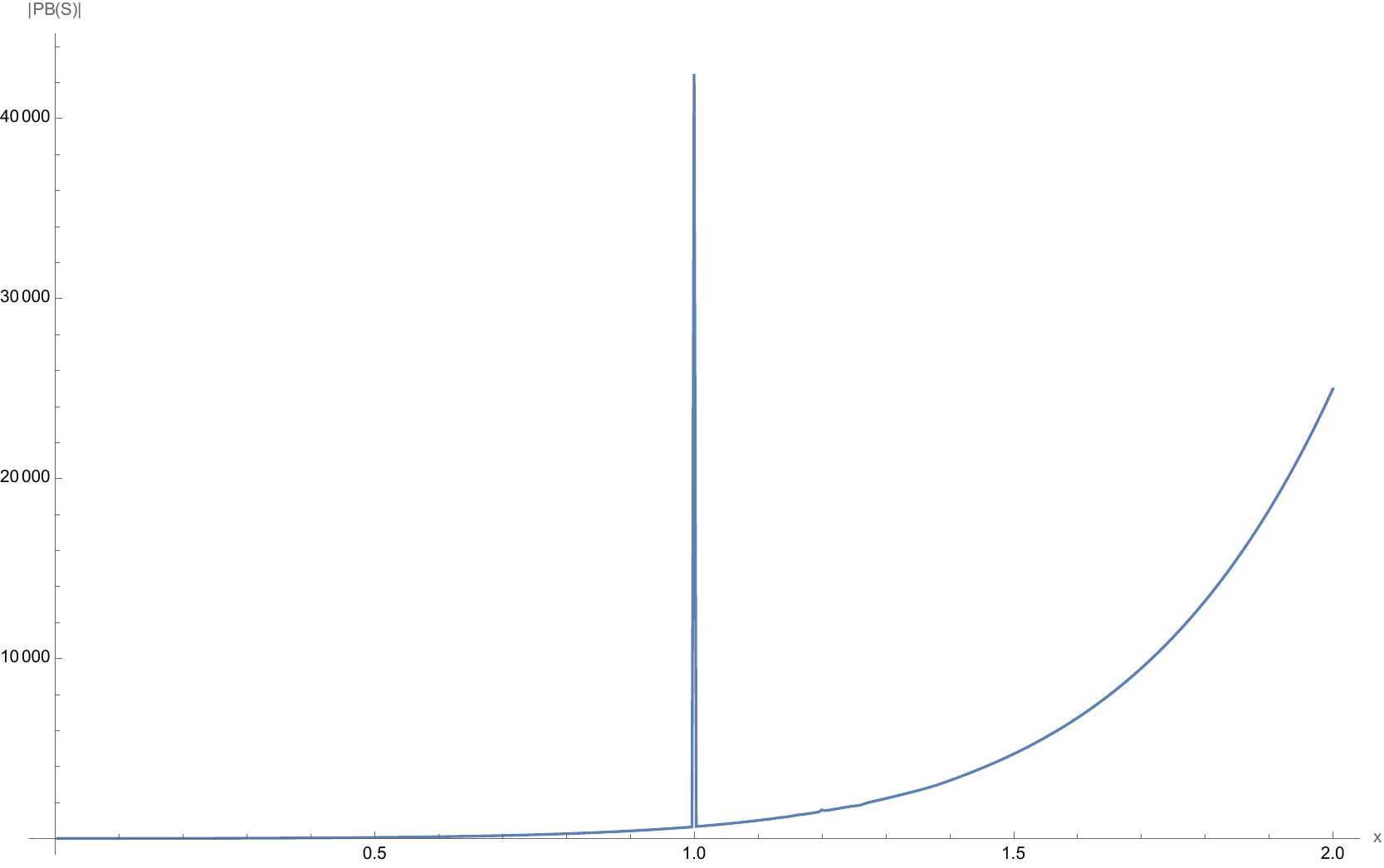}
    \caption{ The plot of $\abs{PB(S)}$ for the Pad\'e-Borel sum of $C_0(p=\pi)$. Here, we plot $\abs{PB(S)}$ along multiple directions given by $S=2\pi i x$, $S=2\pi i(1/2 + ip)x$, and $S=2\pi i(1/2 - ip)x$ direction with $x$ being a positive real number. The poles lies at $x=1$ in each of these directions.  In particular, and the magnitude of the action for these saddles is $S_0=2\pi i$ and $2\pi i(1/2 \pm ip)$ which is the expected behavior for HMW saddles as noted in the text. Note that the strength of the leading saddle $S_0=2\pi i$ is much bigger than the sub-leading ones $S_0 = 2\pi i(1/2 \pm ip)$. For the above figure, we compute the $PB(S)$ by using a $\lfloor 700,700\rfloor$ Pad\'e approximant.}
    \label{fig:C0_p_pade}
\end{figure}

Using the Pad\'e approximants for the the Borel sum $\tilde{B}(S)$, we can identify the location of the singular points in the Borel plane. For the present case of real $p$'s, we denote the Pad\'e-Borel approximant as $\tilde{PB}(S)$. To study its singular points, we plot the function along multiple directions in the complex $S$ plane. We show this in the figure \ref{fig:C0_p_pade} for $p=\pi$.

From the plots, we see that we are able to locate the leading saddle $S_0=2\pi i$ as well as the sub-leading ones $S_0=2\pi i(1/2 \pm ip)$. The sub-leading saddles have a new feature that they can have real and imaginary parts such that the real part is dependent on $p$. 

\section{Analytic structure of DOZZ}
\label{sec:c0_analytic}

In the last section we provided evidence that the perturbative coefficients around the leading saddle point of Liouville theory are Borel summable, and that the resulting Borel sum contains singularities located precisely at the actions of complex solutions of the Liouville equation.  In general, we therefore expect that the DOZZ formula can be written as a trans-series of the form
\begin{equation}
    C_{\rm DOZZ}(\eta_1,\eta_2,\eta_3) = \sum_{N\in\mc{I}}\sum_{n=0}^{\infty} e^{-S_N(\eta_1,\eta_2,\eta_3)/b^2}\, A_{n}^{(N)} b^{2n} 
    \label{eq:cdozztrans}
\end{equation}
The saddle points $N$ appearing in this expression are those indicated in equation (\ref{frodo}).  Of course, the contribution from each individual saddle $N$ in this expression does not make sense on its own, as the corresponding series (the sum over $n$) is asymptotic. 
 The formula only makes sense when interpreted as a trans-series.  In order for this to work, the series expansions around the different saddles must be related in a particular way (see e.g. \cite{Dorigoni:2014hea} for a review).  In particular, they must be related by analytic continuation either in the Borel plane or in the complex $b$ plane.  

To that end, in this section we will study the analytic behavior of $C_0$ as a function of $b^2$.  We will show that the transseries expansion around a saddle has a branch cut running from $b=0$ to $b=\infty$, and that monodromy around this branch cut mixes the perturbative expansion around one saddle with that of a different saddle.  This is an alternate way of seeing how the different perturbative expansions are related to one another.

To begin let us start with the integral expression
\begin{equation}
    \begin{split}
        \log \Gamma_b(x/b) &= \frac{\left( Q/2 - x/b \right)^{2}}{2} \log b  + \int_{0}^{\infty} \frac{dt}{t} \, [\,\, \frac{e^{-xt}}{(1-e^{-t})} \frac{1}{(1-e^{-b^2 t})} - \frac{e^{-t/2}}{(1-e^{-t})}\frac{1}{2\sinh(b^2 t/2)} \\
        &\hspace{6cm} - \frac{\left( Q/2 - x/b \right)^{2}}{2}e^{-t} - \frac{(Q/2 - x/b)}{bt} ]
    \end{split}
\end{equation}
Considered as a function of $b$, this integral is clearly convergent when $b^2>0$.  However, when $b^2<0$ the denominator becomes singular.  This indicates that our expression for $\log \gamma_b(x/b)$, and hence the connected correlator $\log C_0$,   has a branch cut along the negative $b^2$ axis.  

This is not a surprise, as non-analyticities of this sort frequently appear in trans-series expansions.  To understand this better, consider the perturbative expansion as $b\rightarrow0$
\begin{equation}
    \begin{split}
        \log \Gamma_{b}\left(\frac{x}{b} \right) = \frac{\left(\frac{1/2-x}{b} + \frac{b}{2} \right)^2}{2}\log b + \frac{2x-1}{4b^2}\log(2\pi) - \frac{1}{b^2}\int_{1/2}^{x}dt \log \Gamma(t) + C + \sum_{n=0}^{\infty} I_{n} (b^{2})^{2n+1} 
    \end{split}
\end{equation}
So we can view our integral expression, evaluated at $b^2>0$, as capturing the perturbative series around the leading real saddle of Liouville theory.  Considered as a function of complex $b$, the branch cut described above will then instruct us how the perturbative series expanded around different saddle points will mix as we analytically continue $b$.

To proceed, we note that the perturbative expansion described above is actually a Taylor series in $b^4$ and not $b^2$. So life will be simplest if we consider the complex $b^4$ plane, rather than the complex $b$ plane. 
Starting with $b^4>0$, we analytically continue in the complex plane by replacing $b^4$ with $\abs{b^4} e^{i\theta}$ as $\theta$ runs 
from $0$ to $\pi$. 
We now define:
\begin{equation}
    \begin{split}
        F(b^4) &:= b^2\log\left(\frac{\Gamma_b(x/b)}{b^{\frac{1}{2}(Q/2-x/b)^2}} \right) - b^2C \\
    \end{split}
\end{equation}
The above definition is chosen so that the perturbative expansion of the above function is a series in $b^4$ i.e.
\begin{equation}
    F(b^4) = \sum_{n=-1}^{\infty} I_{n} (b^{4})^{n+1}
\end{equation}
in terms of the integral definition, we have
\begin{equation}
    \begin{split}
        F(b^4) &= b^2\int_{0}^{\infty} \frac{dt}{t} \left[  \frac{e^{-xt}}{(1-e^{-t})}\left( \frac{1}{(1-e^{-b^2 t})} - \frac{1}{2} \right) - \frac{e^{-t/2}}{(1-e^{-t})}\frac{1}{2\sinh(b^2 t/2)} \right] \\
        &\hspace{1cm} - \int_{0}^{\infty} \frac{dt}{t} \left[ \frac{(\frac{1}{2} - x) + \frac{b^4}{4} }{2}e^{-t} + \frac{\frac{1}{2} - x}{t} \right] \\
    \end{split}
\end{equation}
where the integrand is slightly modified since we have removed the constant $C$. Then
\begin{equation}
    \begin{split}
        F(b^4) &= b^2\int_{0}^{\infty} \frac{dt}{t} \left[  \frac{\cosh(b^2 t/2) e^{-xt}}{(1-e^{-t})} - \frac{e^{-t/2}}{(1-e^{-t})}\right]\frac{1}{2\sinh(b^2 t/2)} + G(b^4) \\
        &= b^2\int_{0}^{\infty} \frac{dt}{t} \left[  \frac{\cosh(t/2) e^{-xt/b^2}}{(1-e^{-t/b^2})} - \frac{e^{-t/2b^2}}{(1-e^{-t/b^2})}\right]\frac{1}{2\sinh(t/2)} + G(b^4) \\
        &= \sum_{M=0}^{\infty} (b^4)^{1/2} \int_{0}^{\infty} \frac{dt}{t} \left[ \frac{e^t+1}{2} e^{-(x+M)t/b^2} - e^{-(\frac{1}{2}+M)t/b^2}e^{t/2} \right]\frac{1}{e^t - 1} + G(b^4) \\
    \end{split}
\end{equation}
where we have defined $G(b^4)$ as the second line of the previous equation. This term will not be important when we compute the discontinuity or study the analytic structure of $F$, 
but it is important for the convergence of the integral. We now deform $b^4$ to $\abs{b^4}e^{i\theta}$.
To keep the integral convergent we must change the contour of integration appropriately. In particular, we shift $t$ to $\abs{t}e^{i\theta/2}$. We have
\begin{equation}
    \begin{split}
        F(\abs{b^4}e^{i\pi}) &= \sum_{M=0}^{\infty} i\abs{b^4}^{1/2} \int_{0}^{i\infty(1-i\varepsilon)} \frac{dt}{t} \left[ \frac{e^t+1}{2} e^{i(x+M)t/\sqrt{\abs{b^4}}} - e^{(\frac{1}{2}+M)t/\sqrt{\abs{b^4}}}e^{t/2} \right]\frac{1}{e^t - 1} + G(b^4) \\
        &= \sum_{M=0}^{\infty} i\abs{b^4}^{1/2} \int_{0}^{\infty(1-i\varepsilon)} \frac{dT}{T} \left[ \frac{e^{iT}+1}{2} e^{-(x+M)T/\sqrt{\abs{b^4}}} - e^{-(\frac{1}{2}+M)T/\sqrt{\abs{b^4}}}e^{iT/2} \right]\frac{1}{e^{iT} - 1} + G(b^4) \\
    \end{split}
\end{equation}
and
\begin{equation}
    \begin{split}
        F(\abs{b^4}e^{-i\pi}) &= \sum_{M=0}^{\infty} -i\abs{b^4}^{1/2} \int_{0}^{-i\infty(1+i\varepsilon)} \frac{dt}{t} \left[ \frac{e^t+1}{2} e^{-i(x+M)t/\sqrt{\abs{b^4}}} - e^{-i(\frac{1}{2}+M)t/\sqrt{\abs{b^4}}}e^{t/2} \right]\frac{1}{e^t - 1} + G(b^4) \\
        &= \sum_{M=0}^{\infty} -i\abs{b^4}^{1/2} \int_{0}^{\infty(1+i\varepsilon)} \frac{dT}{T} \left[ \frac{e^{-iT}+1}{2} e^{-(x+M)T/\sqrt{\abs{b^4}}} - e^{-(\frac{1}{2}+M)T/\sqrt{\abs{b^4}}}e^{-iT/2} \right]\frac{1}{e^{-iT} - 1} + G(b^4) \\
        &= \sum_{M=0}^{\infty} i\abs{b^4}^{1/2} \int_{0}^{\infty(1+i\varepsilon)} \frac{dT}{T} \left[ \frac{e^{iT}+1}{2} e^{-(x+M)T/\sqrt{\abs{b^4}}} - e^{-(\frac{1}{2}+M)T/\sqrt{\abs{b^4}}}e^{iT/2} \right]\frac{1}{e^{iT}-1} + G(b^4) \\
    \end{split}
\end{equation}
Using this we find a discontinuity in $b^4$ along the negative axis real axis:
\begin{equation}
    \begin{split}
        &F(\abs{b^4}e^{i\pi}) - F(\abs{b^4}e^{-i\pi}) \\
        &\quad = i\abs{b^4}^{1/2}\sum_{M=0}^{\infty} \left( \int_{0}^{\infty(1-i\varepsilon)} - \int_{0}^{\infty(1+i\varepsilon)} \right) \frac{dT}{T} \left[ \frac{e^{iT}+1}{2} e^{-(x+M)T/\sqrt{\abs{b^4}}} - e^{-(\frac{1}{2}+M)T/\sqrt{\abs{b^4}}}e^{iT/2} \right]\frac{1}{e^{iT}-1} \\
        &\quad = i\abs{b^4}^{1/2} (2\pi i)\sum_{M=0}^{\infty} \sum_{N=1}^{\infty} \frac{1}{2\pi i N}\left[ e^{-2\pi N(x+M)/\sqrt{\abs{b^4}}} - (-1)^{N} e^{-2\pi N(\frac{1}{2}+M)/\sqrt{\abs{b^4}}} \right] \\
        &\quad = i\abs{b^4}^{1/2} \sum_{M=0}^{\infty} \log(\frac{1 + e^{-2\pi (\frac{1}{2}+M)/\sqrt{\abs{b^4}}}}{1 - e^{-2\pi (x+M)/\sqrt{\abs{b^4}}}} ) 
    \end{split}
\end{equation}
To get the shift in $\Gamma_b(x/b)$, we exponentiate the above after dividing by a factor of $b^2$, so
\begin{align}
    \label{shiftgammab}
    e^{\frac{1}{b^2}F(\abs{b^4}e^{i\pi})} &\rightarrow e^{\frac{1}{b^2}F(\abs{b^4}e^{-i\pi})} \exp(\sum_{M=0}^{\infty} \log(\frac{1 + e^{-2\pi (\frac{1}{2}+M)/\sqrt{\abs{b^4}}}}{1 - e^{-2\pi (x+M)/\sqrt{\abs{b^4}}}} ) ) \\
    \Gamma_b(x/b) &\rightarrow \Gamma_b(x/b) \prod_{M=0}^{\infty} \left( \frac{1 + e^{-2\pi (\frac{1}{2}+M)/\sqrt{\abs{b^4}}}}{1 - e^{-2\pi (x+M)/\sqrt{\abs{b^4}}}} \right)
\end{align}
We can now apply this to understand how $C_0$ shifts as we perform monodromy in the complex $b$ plane. In the expression for $C_0$, we have 
\begin{equation}
    C_0 \propto \frac{\Gamma_b(\frac{2}{b})}{\Gamma_b(\frac{1}{b})^3}  \frac{ \Gamma_b(\frac{3\eta-1}{b})\, \Gamma_b(\frac{1-\eta}{b})^{3} \, \Gamma_b(\frac{\eta}{b})^{3} \, \Gamma_b(\frac{2-3\eta}{b})}{\Gamma_b(\frac{2\eta}{b})^{3} \, \Gamma_b(\frac{2(1-\eta)}{b})^{3}}
\end{equation}
Due to the equal number of $\Gamma_b$'s in the numerator and denominator, the term independent of $x$ in (\ref{shiftgammab}) will cancel. We conclude that the monodromy $C_0$ across our branch cut will take the form
\begin{equation}
    \begin{split}
        C_0 &\rightarrow C_0 \frac{(1 + e^{-2\pi i(3\eta-1)/b^2})(1 + e^{-2\pi i(1-\eta)/b^2})^3(1 + e^{-2\pi i\eta/b^2})^3(1 + e^{-2\pi i(2-3\eta)/b^2})(1 + e^{-4\pi i/b^2})}{(1 + e^{-4\pi i\eta/b^2})^3(1 + e^{-2\pi i(2-2\eta)/b^2})^3(1 + e^{-2\pi i/b^2})^3} \\
        &\rightarrow C_0 \left( 1 + e^{-2\pi i(3\eta-1)/b^2} + e^{-2\pi i\eta/b^2} + \dots \right) \\
    \end{split}
\end{equation}
In the above we only kept terms with $M=0$. The higher $M$ terms will give more general shifts. Indeed, these are precisely the shifts we expect from the saddle points in HMW. The terms with $M\ne0$ give the more general shifts of the form
\begin{equation}
        C_0 \rightarrow C_0\, e^{2\pi i (n_1 x + n_2)/b^2} \,, \quad n_1, n_2 \in \mathbb{Z}
\end{equation}
This is precisely the mixing that is required in order to render 
the transseries expression for $C_0$, as defined in equation (\ref{eq:cdozztrans}), sensible.

\section*{Acknowledgements}

We are very grateful to E. D'Hoker, G. Dunne, M. Matone, S. Ribault, S. Shenker, M. \"Unsal and E. Witten for useful conversations.  A.M. and V.M. are supported in part by the Natural Sciences and Engineering Research Council of Canada (NSERC), funding reference number SAPIN/00047. N.B. is supported in part by the Sherman Fairchild Foundation and the U.S. Department of Energy, Office of Science, Office of High Energy Physics Award Number DE-SC0011632. S.C. is supported by the U.S. Department of Energy, Office of Science, Office of High Energy Physics of U.S. Department of Energy under grant Contract Number  DE-SC0012567 (High Energy Theory research), DOE Early Career Award  DE-SC0021886, and the Packard Foundation Award in Quantum Black Holes and Quantum Computation.

\appendix

\section{Review of Resurgence}
\label{sec:reviewres}

In this appendix, we review briefly a few aspects of resurgence, Borel resummation of asymptotic series for perturbative expansions, and how they are used to infer the existence of non-perturbative effects.

In many cases it is possible to decode the behavior of non-perturbative effects from the asymptotic behavior of the coefficients in a perturbative expansion. This is the basic idea behind resurgence which by now has been successfully applied to study a variety of quantum systems (see e.g. \cite{tHooft:1977xjm, Dunne:2012ae,
 Marino:2012zq,Dorigoni:2014hea, Jentschura:2004jg, Dunne:2014bca,Fitzpatrick:2016ive,Benjamin:2023uib}).

Let us consider an observable (say the partition function) $Z(\hbar)$ of a quantum mechanical theory, computed as a perturbative series in $\hbar$.
In general we expect an expansion of the form
\begin{equation}
    Z(\hbar) = \int D\phi \, e^{-\frac{S[\phi]}\hbar} \simeq e^{-\frac{S_{\rm cl}}\hbar}\sum_{n=0}^{\infty}a_{n} \hbar^{n+\alpha},
\label{eq:zhbar_pert}
\end{equation}
for some constant $\alpha$. The observable is defined in principle by a path integral over some fields $\phi$, weighted by an action $S[\phi]$.  We have assumed that this observable has a leading saddle with classical action $S_{\rm cl}$, and the coefficients $a_n$ are then computed by expanding the field around this background solution i.e. $\phi = \phi_{\rm cl} + \delta\phi$ and expanding in $\delta\phi$ or $\hbar$. Generically, this series is an asymptotic series rather than a convergent one, with $a_{n}$ growing factorially with $n$. 

We can consider the Borel series $B(S)$ defined as
\begin{equation}
    \label{eq:bs}
    B(S) \equiv \sum_{n=0}^{\infty} \frac{a_{n}}{n!} S^{n},
\end{equation}
where this sum now has a finite radius of convergence (assuming $a_n \sim n!$ at large $n$). If we analytically continue $B(S)$ to the entire complex $S$ plane (known as the Borel plane), then we can define the Borel resummed observable $\tilde{Z}(\hbar)$ as the following:
\begin{equation}
    \tilde{Z}(\hbar) = \frac{1}{\hbar} \int_{0}^{\infty} dS \, B(S) \, e^{-\frac S \hbar}
        \label{eq:zhbar_resummed}
\end{equation}
If we plug (\ref{eq:bs}) into (\ref{eq:zhbar_resummed}) and switch the sum and integral, we get back the perturbative expansion (\ref{eq:zhbar_pert}). In principle, however (\ref{eq:zhbar_resummed}) allows us to go beyond and have a well defined expression for the observable $Z(\hbar)$ at generic values of the coupling constant. This method of integral transform of $B(S)$ can be somewhat ambiguous if there are singularities in the analytically continued function $B(S)$. For example, $B(S)$ may contain a singular point 
at some value $S=S_{0}$ in the Borel plane, 
which by deforming the contour of integration in equation (\ref{eq:zhbar_resummed}) would appear to contribute to $\tilde{Z}(\hbar)$ a term of the form $e^{-S_0/\hbar}$. (The fact that the coefficients $a_n$ grow as $n! S_0^{-n}$ is true even if the singularity $S_0$ in the Borel plane is not a simple pole; see e.g. \cite{Marino:2012zq} for discussion.) 

An intuitive argument as to why a singularity in the Borel plane corresponds to a factorial growth in the coefficients can be given as follows. Let us consider the path integral 
\begin{equation}
Z(\hbar) = \int D\phi~ e^{-S[\phi]/\hbar} \approx \sum_n c_n \hbar^{n} ~.
\end{equation}
We may extract the coefficient $c_n$ from a contour integral
\begin{equation}
c_{n-1} = \int_C d\hbar \int D\phi~ \hbar^{-n} e^{-S[\phi]/\hbar} = \int_C d\hbar \int D\phi~ e^{-S[\phi]/\hbar-n \log \hbar}
\end{equation}
where $C$ circles the origin.
In the large $n$ limit we expect this to be dominated by a saddle point in both the $\hbar$ and $\phi$ integrals, which would obey the saddle point equations
\begin{equation}
\frac{\delta S}{\delta \phi}=0,~~~~~\hbar = \frac Sn
\end{equation}
The first equation says that we are looking for a classical (instanton) solution to the equations of motion -- we will denote the action of this instanton $S_0$.  
Evaluating the integral at the saddle point gives
\begin{equation}
c_{n-1} \approx \left(\frac ne\right)^n \left(S_0\right)^{-n} \approx n! \left(S_0\right)^{-n} \,,\quad  (n\gg 1)
\end{equation}

In general, the behavior of $B(S)$ in the Borel plane can be quite intricate with function having many poles or branch cuts. The leading asymptotic for $a_n$ at large $n$ is determined by the singular point closest to $S=0$. Let us consider the following Borel sum as an example:
\begin{equation}
    B(S) = \sum_{N} \frac{R_N}{1-\frac{S}{S_N}} =  \sum_{n=0}^{\infty} S^n \sum_{N} R_N S_{N}^{-n} 
\end{equation}
where $S_{N}$ are the location of poles in the Borel plane and $N$ is an index that labels them say $(N=0,1,2,\dots)$. We can expect such a Borel sum to arise from perturbative coefficients
\begin{equation}
    a_n = n!\left(\sum_{N} R_N S_{N}^{-n}  \right) = n! \left( R_0 S_{0}^{-n} + R_1 S_{1}^{-n} + \dots  \right) 
\end{equation}
assuming $\abs{S_0} < \abs{S_1} < \abs{S_2} < \dots$, we see that at large $n$, the first term dominates the second one, the second dominates the third and so on. Thus numerically extracting the information about the subleading saddles is exponentially harder. In our analysis of the DOZZ formula, we will encounter a very similar situation. However, we will see that it is still possible to find them using the Pad\'e approximants. The exact details are given in section \ref{sec:res_dozz} when we use them explicitly.

As already mentioned above, in applying these arguments to general path integrals, one crucial question is whether the singularities in $B(S)$ lie on the positive real axis which can lead to ambiguous results \cite{Dunne:2016jsr}. Sometimes the ambiguity can be due a physical effects like decays in unstable potentials or unphysical in which case they need to be canceled by other non-perturbative terms. This kind of phenomena has been studied in several systems in quantum mechanics \cite{Dunne:2014bca, Lipatov:1976ny, Balitsky:1985in, Dunne:2013ada}. In these examples, one can write well defined and unambiguous by including all non-perturbative effects. In more complicated cases like in quantum field theories, however, one can only constrain the nature of non-perturbative effects by studying the perturbative coefficients $a_n$ and the singularities in $B(S)$ in the Borel plane.

\section{Resurgence in Gamma Function}
\label{appGamma}
Let us start with the integral definition of the Gamma function
\begin{equation}
    \Gamma(z) = \int_{0}^{\infty} dt \, t^{z-1} e^{-t}
\end{equation}
The above integral converges for $\Re z >0$. Since this is an one-dimensional integral, we can explicitly write this in the form of a Borel transform by a simple change of variables. Consider $t=ze^{\phi}$
\begin{equation}
    \label{gamma_phi}
    \begin{split}
        \Gamma(z) &= z^{z} \int_{-\infty}^{\infty} d\phi \, e^{-z(e^{\phi} - \phi)} \\
        &= z^{z} \int_{-\infty}^{0} d\phi \, e^{-z(e^{\phi} - \phi)} + z^{z} \int_{0}^{\infty} d\phi \, e^{-z(e^{\phi} - \phi)} \\
    \end{split}
\end{equation}
We now rewrite these expressions as a Borel Transform by defining $S = e^{\phi}-\phi-1$:
\begin{equation}
    \frac{ \Gamma(z)}{z^z e^{-z}} = \int_{0}^{\infty} dS \, e^{-zS} \left[  \frac{1}{e^{\phi_{+}(s)}-1} - \frac{1}{e^{\phi_{-}(s)}-1} \right] = \int_{0}^{\infty} dS \, e^{-zS} \, B(S)
\end{equation}
where $\phi_{\pm}(S)$ indicates the different branch choice when we invert the function $S = e^{\phi}-\phi-1$ to express $\phi$ in terms of $S$ implicitly. In this case, we then have the Borel sum to be
\begin{equation}
    B(S) = \frac{1}{e^{\phi_{+}(S)}-1} - \frac{1}{e^{\phi_{-}(S)}-1}
\end{equation}
It is interesting to look at the two limits $S\rightarrow0$ and $S\rightarrow\infty$. In the first limit, we have
\begin{equation}
    \begin{split}
        \phi^2 = 2S \implies \phi_{\pm} \approx \pm \sqrt{2S} \implies B(S) \approx \sqrt{\frac{2}{S}} \,, \quad (S\rightarrow0)
    \end{split}
\end{equation}
Plugging this back in the integral gives the expected `one-loop' part of the Stirling formula $\sqrt{2\pi/z}$. In the second limit, we have
\begin{equation}
    \begin{split}
        \exp(\phi_{+}) &\approx S \,,\quad \phi_{-} \approx -S \implies B(S) \approx \frac{1}{S-1} - \frac{1}{e^{-S}-1} \\
        B(S) &= \left(\frac{1}{S} + \frac{1}{S^2} + \dots \right) + \left(1 + e^{-S} + e^{-2S} + \dots \right) \,, \quad (S\rightarrow\infty)
    \end{split}
\end{equation}
This shows that we can have a non-trivial behavior of the Borel sum in the small and large $S$ limits.  Now, let us look at the singular points in the Borel plane. From the above expression, we can see that the Borel sum $B(S)$ has a singular point when $\phi_{\pm} = 2\pi i N$ for $N\in \mathbb{Z}$. Close to this point $\phi = 2\pi iN + x$, we have
\begin{equation}
    e^x - 2\pi i N - x - 1  = S \implies x_{\pm} \approx \pm \sqrt{2(S+2\pi i N)}  
\end{equation}
\begin{equation}
    B(S) \approx \sum_N \frac{\sqrt{2}}{\sqrt{S-2\pi i N}}(1 + \mathcal{O}(S-2\pi i N)) 
\end{equation}
These singular are directly related to the saddle points in the integral definition. From (\ref{gamma_phi}), the saddle point equation for $z\rightarrow\infty$ is
\begin{equation}
    e^{\phi} = 1 \implies \phi = 2\pi i N \,, \quad N\in \mathbb{Z}
\end{equation}
Substituting this in the integral definition, they lead to non-perturbative terms of the form $e^{2\pi iN z}$.

\subsection*{Perturbative expansion}
\begin{figure}[t!]
    \centering
    \includegraphics[scale=0.5]{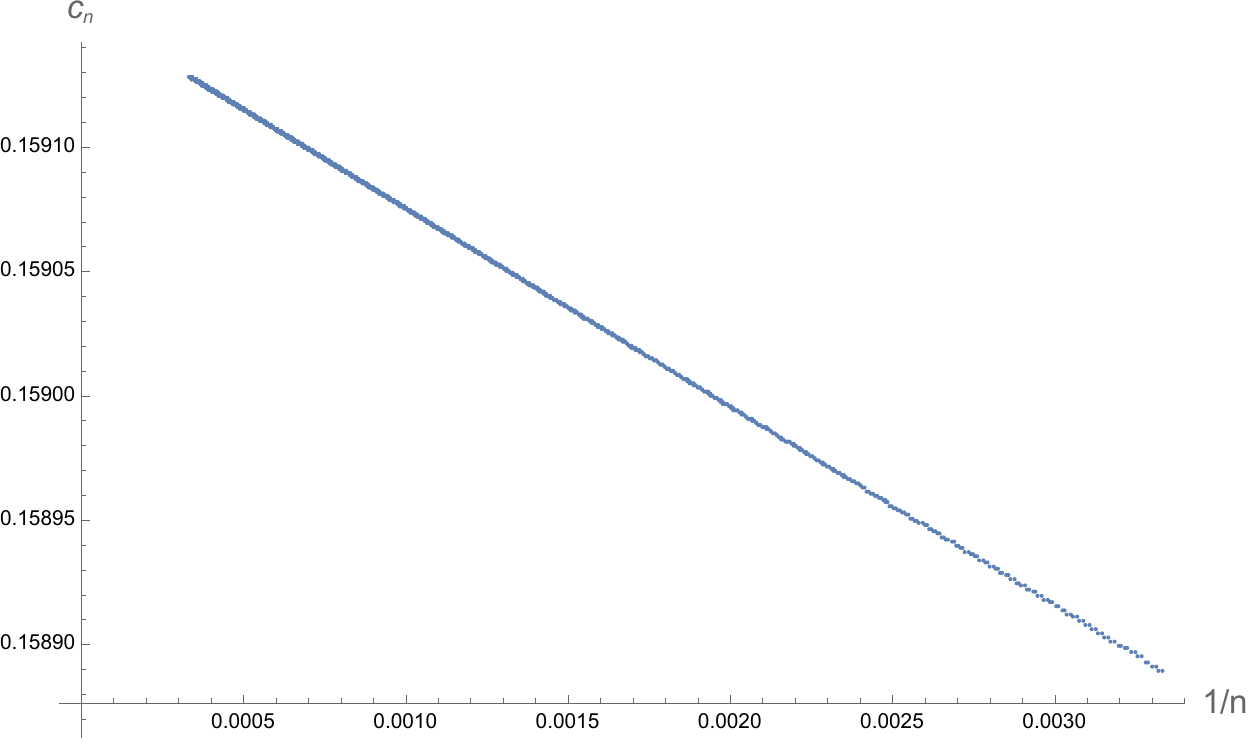}
    \caption{The plot of $c_n$ for the Gamma function. We compute the $c_n$ upto $n=3000$. The straight line fit $y=0.159155 - 0.0796905\, x$ can be used to extract the values of $\abs{S_0}$ and $\alpha$. Using this, we get $\abs{S_0} = 6.283183$ and $\alpha = 0.499290$.}
    \label{fig:gamma_cn}
\end{figure}

We will now compute the perturbative expansion of the Gamma function in the limit of $z\rightarrow\infty$. Although, we can do this computation directly from the integral definition by expanding around saddle point, we will follow steps which are similar to the case of DOZZ formula. We start with the integral definition of the $\log \Gamma(z)$.
\begin{equation}
    \log \Gamma(z) = \int_{0}^{\infty} \frac{dt}{t} \, \left[ \frac{e^{-zt}-e^{-t}}{1-e^{-t}} + (z-1)e^{-t} \right]
\end{equation}
using this expression, we can write the perturbative expansion of $\log \Gamma(z)$ as
\begin{equation}
    \log \frac{\Gamma(z)}{z^{z} e^{-z}} = -\frac{1}{2}\log\frac{z}{2\pi} + \sum_{n=0}^{\infty} \frac{a_n}{z^{2n+1}} \,, \quad (z\rightarrow\infty)
\end{equation}
where the coefficients $a_n$ are given in terms of the Bernoulli numbers
\begin{equation}
    a_n = \frac{B_{2n+2}}{(2n+1)(2n+2)} \,, \quad B_{2n} = \frac{(-1)^{n+1}2(2n)!}{(2\pi)^{2n}} \zeta(2n)
\end{equation}
\begin{equation}
    a_n = 2(-1)^n (2n)! \frac{\zeta(2n+2)}{(2\pi)^{2n+2}} = -2 (2n)! \left( \frac{1}{(2\pi i)^{2n+2}} + \frac{1}{(4\pi i)^{2n+2}} + \dots \right)
\end{equation}
comparing these expressions with generic form of these coefficients (\ref{gen_an}), we can see that the singularities in the Borel plane for the $\log\Gamma(z)$ are located at $2\pi i N$.
Computing then the perturbative expansion of the Gamma function:
\begin{equation}
    \Gamma(z) = z^{z} e^{-z} \sqrt{\frac{2\pi}{z}} \exp\left(\sum_{n=0}^{\infty} \frac{a_n}{z^{2n+1}} \right) = z^{z} e^{-z} \sqrt{\frac{2\pi}{z}}  \sum_{n=0}^{\infty} \frac{b_n}{z^n} \,, \quad (z\rightarrow\infty)
\end{equation}
As for the case of DOZZ, the $b_n$ grow as $n!$ at large $n$. 
Due to this we can also expect the location of the singularities to be at $2\pi i N$ in the Borel plane. 
We evaluate the coefficients $b_n$ numerically and study it's behavior at large $n$. Defining
\begin{equation}
    B(S) = \sum_{n=0}^{\infty} d_n S^{n-1/2} \,, \quad d_n = \frac{\sqrt{2\pi} \, b_n}{\Gamma(n+1/2)}
\end{equation}
we can easily check that this reproduces the perturbative expansion for the Gamma function once we substitute it in the integral
\begin{equation}
    \frac{\Gamma(z)}{z^z e^{-z}} = \int_{0}^{\infty} dS \, e^{-zS} \, B(S)
\end{equation}
To compute the value of $\abs{S_0}$ and $\alpha$, we plot the $c_n$'s for large $n$ and fit a straight line to it. This is shown in the figure \ref{fig:gamma_cn}. We find
\begin{equation}
    \boxed{\abs{S_0} = 6.283183 \,, \quad \alpha = 0.499290}
\end{equation}
We can see clearly that the value of the action $\abs{S_0}$ agrees very well with the magnitude of the subleading saddle i.e. $\abs{2\pi i } = 6.283185$. The nature of the singularity is also determined to a high accuracy since to be a square root for which $\alpha=0.5$.

\subsection*{Pad\'e-Borel}

\begin{figure}[t!]
    \centering
    \includegraphics[scale=0.5]{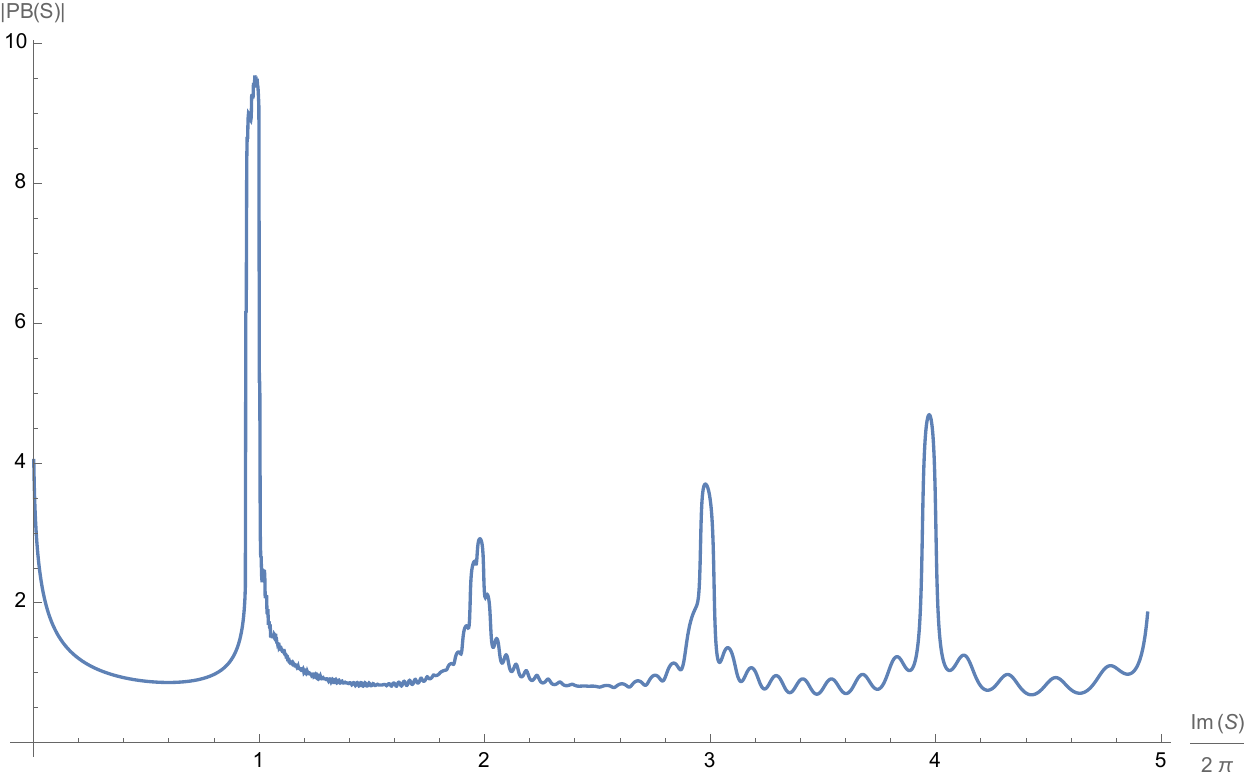}
    \caption{The plot of $\abs{PB(S)}$ for the Pad\'e-Borel sum of the Gamma function. Here, we plot $\abs{PB(S)}$ along the imaginary $S$ axis. We can see that the singularities are located precisely at the points $2\pi i N$. For the above figure, we compute the $PB(S)$ by using a $\lfloor 300,300\rfloor$ Pad\'e approximant.}
    \label{gamma_pade_1}
\end{figure}

\begin{figure}[t!]
    \centering
    \includegraphics[scale=0.35]{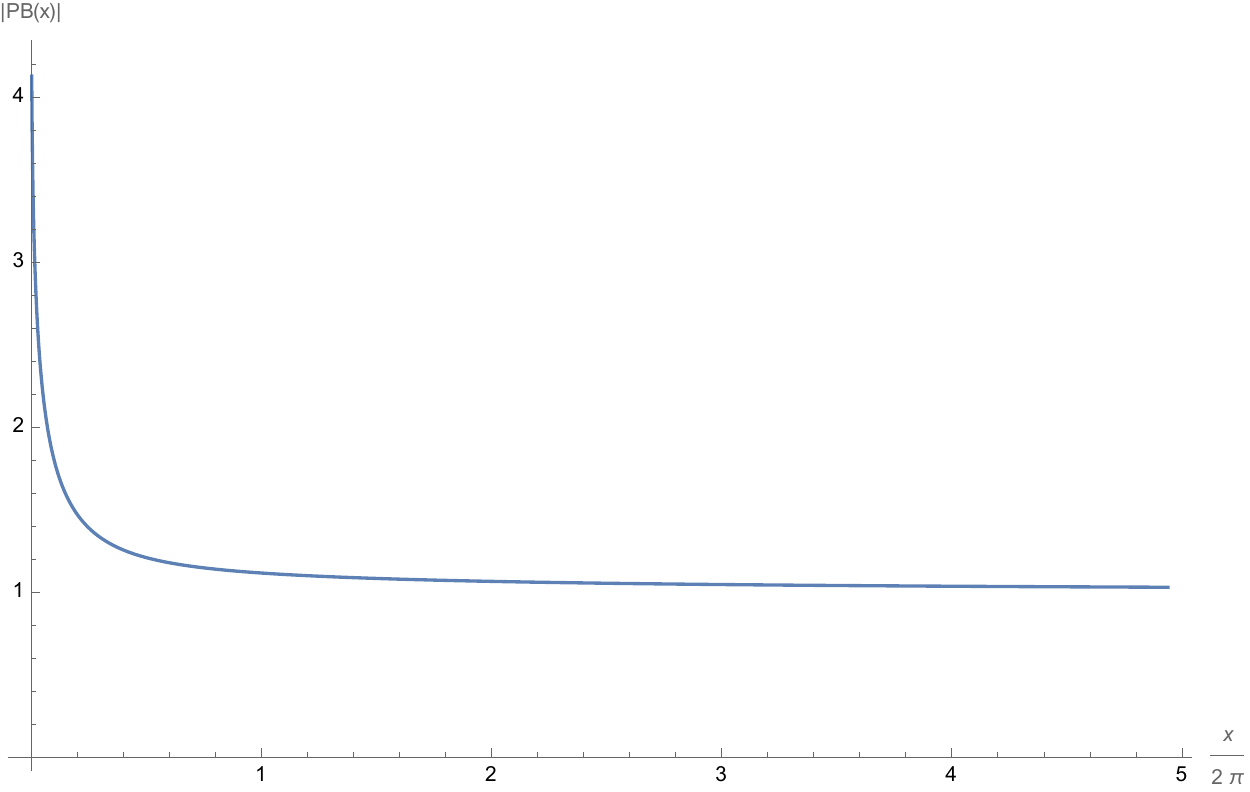}
    \hspace{1cm}
    \includegraphics[scale=0.35]{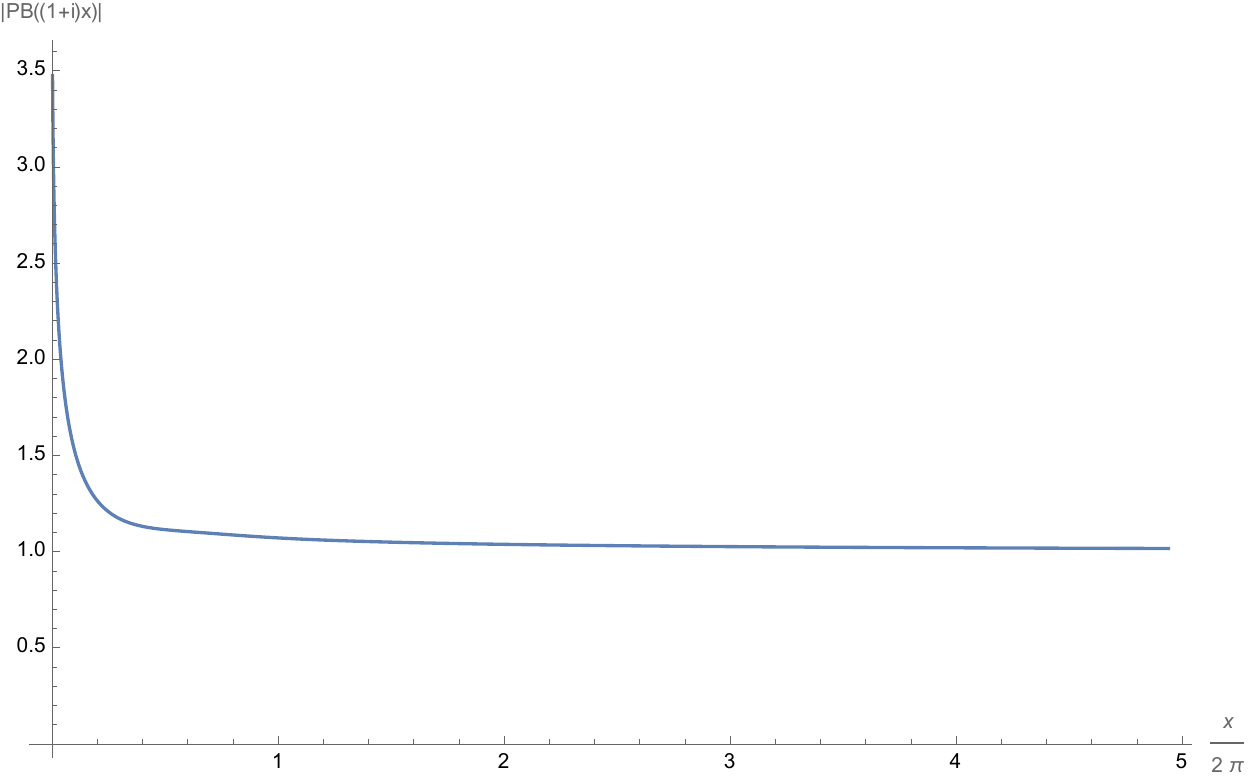}
    \caption{The plot of $\abs{PB(S)}$ for the Pad\'e-Borel sum of the Gamma function. Here, we plot $\abs{PB(S)}$ along $S=x$ and $S=(1+i)x$ direction with $x$ being a positive real number. We see that there is singular behavior along these directions. For the above figure, we compute the $PB(S)$ by using a $\lfloor 300,300\rfloor$ Pad\'e approximant.}
    \label{gamma_pade_2}
\end{figure}

We will now use this Pad\'e approximants to approximate the Borel sum of the Gamma function as a rational function. Using this we can study the analytic behavior of $B(S)$ in the Borel plane. We write
\begin{equation}
    B(S) = \sum_{n=0}^{N} d_n S^{n-1/2} = PB(S) \,,\quad PB(S) = \frac{P_{[N/2]}(S)}{Q_{[N/2]}(S)}
\end{equation}
here, $Q_{\lfloor N/2 \rfloor}$ is a polynomial in $S^{1/2}$ since we have a series expansion in $S^{1/2}$. We can plot this function as a function of the complex $S$ variable and study its behavior. In particular, we find that $PB(S)$ has singular points along the imaginary axis and nowhere else. This is shown in figure \ref{gamma_pade_1}, \ref{gamma_pade_2} where we plot $\abs{PB(S)}$ along different lines passing through the complex $S$-plane. From the figure, we see that the poles lies precisely at the locations $2\pi iN$. The blow up at the origin is due to the $1/\sqrt{S}$ divergence as we take $S\rightarrow0$.

\subsection*{Non-perturbative effects from discontinuity computation}
We will now give analytic argument to observe the non-perturbative saddles in the asymptotic expansion of the Gamma function. This is done by studying the behavior of function directly as function of the `coupling'$(=1/z)$. In particular we are interested in computing $\Gamma(ze^{2\pi i})$ and comparing it with $\Gamma(z)$. Now, we know that the $\Gamma(z)$ has only simple poles and no non-analytic behavior in the complex plane. However, it is interesting to a study a closely related object defined as
\begin{equation}
    F(z) := \frac{\Gamma(z)}{z^z e^{-z}}
\end{equation}
The motivation to consider is as follows: the perturbative expansion and the corresponding Borel transform we wrote earlier was indeed for the function $F(z)$ and it is also more closely related to how the $\Gamma_b(x/b)$ functions appear in the DOZZ formula. Now, it is easy to see that
\begin{equation}
    F(ze^{2\pi i}) = \frac{\Gamma(ze^{2\pi i})}{(e^{2\pi i}z)^{ze^{2\pi i}} e^{-ze^{2\pi i}}} = e^{-2\pi i z}F(z)
\end{equation}
We see that the effect of going around the origin in the complex $z$-plane is take us from the expansion around one saddle to an expansion around another. The exponent in the prefactor in above is precisely the of action of the complex saddle of the integral. We will now derive this same result using a more indirect step by using the integral definition of the $\log\Gamma(z)$ function. Again, this will be more closely related to how the computation works for the DOZZ formula since the $\Gamma_b(x/b)$ have a more complicated behavior in terms of $b$ then $\Gamma_b(z)$.

Starting from the integral definition of the $\log\Gamma$, we can manipulate to get an integral representation for $\log F(z)$
\begin{equation}
    \begin{split}
        \log \Gamma(z) &= \int_{0}^{\infty} \frac{dt}{t} \, \left[ \frac{e^{-zt}-e^{-t}}{1-e^{-t}} + (z-1)e^{-t} \right] \\
        \log F(z) &= -\frac{1}{2}\log \frac{z}{2\pi} + \int_{0}^{\infty}\frac{dt}{t} e^{-zt} \left( \frac{1}{e^t - 1} - \frac{1}{t} + \frac{1}{2} \right) \\
    \end{split}
\end{equation}
we want consider how this expression changes as we move around in the complex $z$ plane. In particular, let us replace $z\rightarrow\abs{z}e^{i\theta}$ can consider changing $\theta$ from $0$ to $2\pi$. It is easy to see that the first term gives us a simple shift of $-i\pi$. The second term with the integral is the more interesting one. To keep the integral convergent, we need to rotate the $t$ contour as $t\rightarrow\abs{t}e^{-i\theta}$. The integrand above has poles at $t=2\pi iN$, where $N\in\mathbb{Z}_{\ne0}$. As we go around the origin in the $z$-plane we get
\begin{equation}
    \begin{split}
    \log F(ze^{2\pi i}) &= \log F(z) - \pi i + (-2\pi i) \sum_{N\in\mathbb{Z}_{\ne0}} \frac{e^{-2\pi izN}}{2\pi iN} \\
    &= \log F(z)  - \pi i - \sum_{N\in\mathbb{Z}_{\ne0}} \frac{e^{-2\pi izN}}{N} \\
    &= \log F(z)  - \pi i + \log(1-e^{-2\pi iz}) - \log(1-e^{2\pi iz}) \\
    &= \log F(z) - \pi i + \log(-e^{-2\pi i z}) \\
    &= \log F(z)  - 2\pi i z + 2\pi ik \,, \quad (k\in\mathbb{Z}) \\
    \end{split}
\end{equation}
where in the last equation the integer $k$ comes from taking the logarithm of $-1$ and the value of $k$ chooses the branch. We can now compute how the function $F(z)$ changes go around the origin in the complex $z$-plane by taking an exponential of the above equation.
\begin{equation}
    F(ze^{2\pi i}) =  e^{-2\pi i z}F(z)
\end{equation}
which is precisely what we found from our previous argument by using the analytic property of the $\Gamma(z)$ function directly.

\section{Asymptotics of \texorpdfstring{$\Gamma_b(x/b)$}{Gammaxb}}
\label{app:gammab}
In this appendix, we derive the asymptotic expansion of the $\Gamma_b(x/b)$ in the semiclassical limit of $b\rightarrow0$ as stated in (\ref{logGammab}). To do this we start with the integral definition: 
\begin{equation}
    \begin{split}
        \log \Gamma_{b}\left(\frac{x}{b} \right) &= \int_{0}^{\infty} \frac{dt}{t} \left[ \frac{e^{-xt/b} - e^{-Qt/2}}{(1-e^{-bt})(1-e^{-t/b})} - \frac{\left( Q/2 - x/b \right)^{2}}{2}e^{-t} - \frac{(Q/2 - x/b)}{t}  \right] \\
        &= \int_{0}^{\infty} \frac{dt}{t} \left[ \frac{e^{-xt} - e^{-t/2 - b^2 t/2}}{(1-e^{-b^2 t})(1-e^{-t})} - \frac{\left( Q/2 - x/b \right)^{2}}{2}e^{-bt} - \frac{(Q/2 - x/b)}{bt}  \right] \\  
        &= \frac{\left( Q/2 - x/b \right)^{2}}{2} \log b  + \int_{0}^{\infty} \frac{dt}{t} \, \bigg[\,\, \frac{e^{-xt}}{(1-e^{-t})} \frac{1}{(1-e^{-b^2 t})} - \frac{e^{-t/2}}{(1-e^{-t})}\frac{1}{2\sinh(b^2 t/2)} \\
        &\hspace{6cm} - \frac{\left( Q/2 - x/b \right)^{2}}{2}e^{-t} - \frac{(Q/2 - x/b)}{bt} \bigg]\\  
    \end{split}
\end{equation}
where we first scaled $t\rightarrow bt$ and then used the identity
\begin{equation}
    \log b  = \int_{0}^{\infty} \frac{dt}{t}\, ( e^{-t} - e^{-bt} )
\end{equation}
We then expand the integrand as we take $b^2 \rightarrow 0$.
\begin{equation}
    \begin{split}
        \log \Gamma_{b}\left(\frac{x}{b} \right) &= \frac{1}{2} \left( \frac{1}{b^2}(1/2 - x)^2 + (1/2 - x) + \frac{b^2}{4} \right)\log b + b^{-2}I_{-1} + I_{0} + b^2 I_{1} + \sum_{n=1}^{\infty} (b^{2})^{2n+1} I_{2n+1} \\
    \end{split}
\end{equation}
where $I_k$ are defined by the integrals
\begin{align}
    I_{-1}(x) &= \int_{0}^{\infty} \frac{dt}{t} \, \left[\frac{e^{-t x}}{\left(1 - e^{-t}\right) t}-\frac{e^{-t/2}}{\left(1-e^{-t}\right) t} -\frac{e^{-t}}{2}(1/2-x)^2  - \frac{1/2-x}{t} \right] \\
    I_{0}(x) &= \int_{0}^{\infty} \frac{dt}{t} \, \left[\frac{e^{-t x}}{2 \left(1 - e^{-t}\right)} - \frac{e^{-t}}{2}\left(1/2-x\right)-\frac{1}{2 t}\right] 
\end{align}
We can evaluate $I_{-1}$ and $I_{0}$ using the following trick. We first observe
\begin{equation}
    \frac{d I_{-1}(x)}{dx} = -2I_{0}(x) \,, \quad I_{-1}(1/2) = 0
\end{equation}
so, evaluating $I_{0}(x)$ fixes $I_{-1}(x)$. Now,
\begin{equation}
    \begin{split}
        2I_{0}(x) &= \int_{0}^{\infty} \frac{dt}{t} \, \left[\frac{e^{-t x}}{\left(1 - e^{-t}\right)} + e^{-t}\left(x-1/2\right)-\frac{1}{t}\right] \\
        &= \log \Gamma(x) + \int_{0}^{\infty} \frac{dt}{t} \, \left[\frac{1}{e^{t} - 1} + \frac{e^{-t}}{2} -\frac{1}{t}\right] \\
        &= \log \Gamma(x) - \frac{1}{2}\log(2\pi) \\        
    \end{split}
\end{equation}
Integrating this expression, we evaluate $I_{-1}(x)$.
\begin{equation}
    I_{-1}(x) = -\int_{1/2}^{x}dt \, \log \Gamma(t) + \frac{2x-1}{4}\log(2\pi) 
\end{equation}
The higher order terms $I_n(x)$, can then be computed easily using integral definitions for the polygamma functions.
\begin{equation}
    \begin{split}
    I_1 &= \int_{0}^{\infty} \frac{dt}{t} \, \left[ \frac{t e^{-t x}}{12 \left(1-e^{-t}\right)}+\frac{e^{-\frac{t}{2}} t}{24 \left(1-e^{-t}\right)}-\frac{e^{-t}}{8} \right] \\
    &= -\left(\frac{1}{12} \frac{d}{dz}\log\Gamma(z) |_{z=x} + \frac{1}{24} \frac{d}{dz}\log\Gamma(z) |_{z=1/2}  \right)  \\
    &= -\left(\frac{\psi(x)}{12} + \frac{\psi(1/2)}{24}   \right)  \\
    \end{split}
\end{equation}
\begin{equation}
    \begin{split}
    I_{2n+1} &= \frac{B_{2n+2}}{(2n+2)!} \left[ \int_{0}^{\infty} dt \,\, \frac{t^{2n}  e^{-xt} }{(1-e^{-t})} + \left( 1 - \frac{1}{2^{2n+1}} \right)  \int_{0}^{\infty} dt \,\,\frac{t^{2n} e^{-t/2}}{(1-e^{-t})} \right] \quad (n \ge 1) \\
    \\
    &=  -\frac{B_{2n+2}}{(2n+2)!}\left[ \psi^{(2n)}(x) + ( 1 - 2^{-(2n+1)}) \psi^{(2n)}(1/2)\right]  \\
    \\
    &=  \frac{B_{2n+2}}{(2n+2)!} (2n)! \left( \zeta(2n+1,x) + ( 1 - 2^{-(2n+1)}) \zeta(2n+1,1/2) \right)\\
    \end{split}
\end{equation}

\bibliography{refs}
\bibliographystyle{JHEP}

\end{document}